\documentclass[12pt,a4paper]{article} 
\usepackage{cite}
\usepackage{graphicx}
\usepackage{amssymb}
\usepackage{a4wide} 

\newcommand{\address}[1]{ \par {\centering #1 \vspace{1.4em} \noindent\par} }

\newcommand*{\re}{\mathop{\mathrm{Re}}} 
\newcommand*{\tr}{\mathop{\mathrm{Tr}}}

\newcommand*{\erfc}{\mathop{\mathrm{erfc}}} 
\newcommand*{\dd}{\mathrm{d}}

\begin{document}

\title{Time problem in quantum mechanics and its analysis by the concept of
weak measurements}

\author{J. Ruseckas and B. Kaulakys}

\maketitle

\address{Institute of Theoretical Physics and Astronomy, Vilnius
University,\\ A. Go\v{s}tauto 12, LT-01108 Vilnius, Lithuania}

\begin{abstract}
The model of weak measurements is applied to various
problems, related to the time problem in quantum mechanics. The review and
generalization of the theoretical analysis of the time problem in quantum
mechanics based on the concept of weak measurements are presented. 
A question of the time interval the system
spends in the specified state, when the final state of the system is given, is
raised. Using the concept of weak measurements the expression for such time is obtained. The
results are applied to the tunneling problem. A procedure for the calculation
of the asymptotic tunneling and reflection times is proposed. Examples for
$\delta$-form and rectangular barrier illustrate the obtained results. Using the
concept of weak measurements the arrival time probability distribution is
defined by analogy with the classical mechanics. The proposed procedure is
suitable to the free particles and to particles subjected to an external
potential, as well. It is shown that such an approach imposes an inherent
limitation to the accuracy of the arrival time definition.
\end{abstract}

\section{Introduction}

The time plays a special role in quantum mechanics. Unlike other observables, time
remains a classical variable. It cannot be simply quantized because, as it is
well known, the self-adjoint operator of time does not exist for the bounded
Hamiltonians. The problems related to time also arise from the fact that in quantum
mechanics many quantities cannot have definite values simultaneously. The
absence of the time operator makes this problem even more complicated. However,
in practice the time is often important for an experimenter. If quantum
mechanics can correctly describe the outcomes of the experiments, it must also
give the method for the calculation of the time the particle spends in some
region.

The most-known problem of time in quantum mechanics is the so-called ''tunneling
time problem''. Tunneling phenomena are inherent in numerous quantum systems,
ranging from an atom and condensed matter to quantum fields. There have been many
attempts to define a physical time for tunneling processes, since this question
has been raised by MacColl \cite{maccol} in 1932. This question is still the
subject of much controversy, since numerous theories contradict each other in
their predictions for {}``the tunneling time''. Some of these theories predict
that the tunneling process is faster than light, whereas the others state that
it should be subluminal. This subject has been covered in a number of reviews
(Hauge and St\o vneng \cite{haugestovneng}, 1989; Olkholovsky and Recami
\cite{olkhovskyrecami}, 1992; Landauer and Martin \cite{landauermartin}, 1994
and Chiao and Steinberg \cite{chiao}, 1997). The fact that there is a time
related to the tunneling process has been observed experimentally
\cite{gueret1,gueret2,esteve,enders,ranfagni,spielmann,heitmann,balcou,garrison,martinez}.
However, the results of the experiments are ambiguous.

Many problems with time in quantum mechanics arise from the noncommutativity of
the operators. The noncommutativity of the operators in
quantum mechanics can be circumvented by using the concept of weak measurements.
The concept of weak measurement was proposed by Aharonov, Albert and Vaidman
\cite{Aharonov2,Aharonov1,Duck,Aharonov3,Aharonov4,Aharonov5}. Such an approach
has several advantages. It gives, in principle, the procedure for measuring the
physical quantity. Second, since in the classical mechanics all quantities can have
definite values simultaneously, weak measurements give the correct classical
limit. The concept of weak measurements has been already applied to the time
problem in quantum mechanics \cite{Ruseckas1,Ruseckas2,Ruseckas5}.

The time in classical mechanics describes not a single state of the system but
the process of the evolution. This property is an essential concept of the time.
We speak about the time belonging to a certain evolution of the system. If the
measurement of the time disturbs the evolution we cannot attribute this measured
duration to the undisturbed evolution. Therefore, we should require that the
measurement of the time does not disturb the motion of the system. This means
that the interaction of the system with the measuring device must be
asymptotically weak. In
quantum mechanics this means that we cannot use the strong measurements
described by the von-Neumann's projection postulate. We have to use the weak
measurements of Aharonov, Albert and Vaidman
\cite{Aharonov2,Aharonov1,Duck,Aharonov3,Aharonov4,Aharonov5}, instead.

We proceed as follows: In Sec.~\ref{sec:concept} we present the model of the
weak measurements. Sec.~\ref{sec:The-time-on} presents the time on condition
that the system is in the given final state. In Sec.~\ref{sec:Tunneling-time},
our formalism is applied to the tunneling time problem. In
Sec.~\ref{sec:Weak-measurement-of} the weak measurement of the quantum arrival
time distribution is presented. Section \ref{sec:concl} summarizes our findings.

\section{\label{sec:concept}The concept of weak measurements}

In this section we present the concept of weak measurement, proposed by
Aharonov, Albert and Vaidman
\cite{Aharonov2,Aharonov1,Duck,Aharonov3,Aharonov4,Aharonov5}. We measure
quantity represented by the operator $\hat{A}$.

We have the detector in the initial state $|\Phi\rangle$. For a weak
measurement to provide the meaningful information the measurements must be
performed on an ensemble of identical systems. It is supposed that each system with its own detector
is prepared in the same initial state.  After measurement the readings of the detectors are
collected and averaged.

Our model consists of the system \textbf{S} under consideration and of the
detector \textbf{D}. The total Hamiltonian is 
\begin{equation}
\hat{H}=\hat{H}_{\mathrm{S}}+\hat{H}_{\mathrm{D}}+\hat{H}_{\mathrm{I}}
\end{equation}
 where $\hat{H}_{\mathrm{S}}$ and $\hat{H}_{\mathrm{D}}$ are the Hamiltonians of
the system and  detector, respectively. We take the operator describing
the interaction between the particle and the detector of the form
\cite{Ruseckas1,Ruseckas2,Ruseckas3,Ruseckas6,Ruseckas7,Ruseckas8}
\begin{equation}
\hat{H}_I=\lambda\hat{q}\hat{A},\label{eq:sec2:11a}
\end{equation}
where $\lambda$ characterizes the strength of the interaction between the system
and detector. The small parameter $\lambda$ ensures the undisturbance of the system's
evolution. The measurement duration is $\tau$. In this section we assume that the
interaction strength $\lambda$ and the time $\tau$ are small. The operator
$\hat{q}$ acts in the Hilbert space of the detector. We require the
spectrum of the operator $\hat{q}$ to be continuous. For simplicity, we can consider this
operator to be the coordinate of the detector. The momentum which conjugate to $q$ is
$p_q$.

The interaction operator (\ref{eq:sec2:11a}) only slightly differs from the one
used by Aharonov, Albert and Vaidman \cite{Aharonov1}. The similar interaction
operator has been considered by von Neumann \cite{vNeum} and has been widely
used in the strong measurement models (e.g.,
\cite{Ruseckas8,Aharonov6,joos,caves,milb,gagen} and many others).

Hamiltonian~(\ref{eq:sec2:11a}) represents a constant force acting on the
detector. This force results in the change of momentum of the detector. From the
classical point of view, the change of the momentum is proportional to the force
acting on the detector. Since interaction strength $\lambda$ and the duration of
the measurement $\tau$ are small, the average $\langle\hat{A}\rangle$ should not
change significantly during the measurement. The action of the Hamiltonian
(\ref{eq:sec2:11a}) results in the small change of the mean detector momentum
$\langle\hat{p}_q\rangle-\langle\hat{p}_q\rangle_0=-\lambda\tau\langle\hat{
A}\rangle$, where
$\langle\hat{p}_q\rangle_0=\langle\Phi(0)|\hat{p}_q|\Phi(0)\rangle$ is the mean
momentum of the detector at the beginning of the measurement and
$\langle\hat{p}_q\rangle=\langle\Phi(\tau)|\hat{p}_q|\Phi(\tau)\rangle$ is the
mean momentum of the detector after the measurement. Therefore, in analogy to
Ref.~\cite{Aharonov1}, we define the {}``weak value'' of the average
$\langle\hat{A}\rangle$, 
\begin{equation}
\langle\hat{A}\rangle=\frac{\langle\hat{p}_q\rangle_0-\langle\hat{p}_q\rangle}{
\lambda\tau}.\label{eq:sec2:defin}
\end{equation}

At the moment $t=0$ the density matrix of the whole system is
$\hat{\rho}(0)=\hat{\rho}_{\mathrm{S}}(0)\otimes\hat{\rho}_{\mathrm{D}}(0)$,
where $\hat{\rho}_{\mathrm{S}}(0)$ is the density matrix of the system and
$\hat{\rho}_{\mathrm{D}}(0)=|\Phi\rangle\langle\Phi|$ is the density matrix of
the detector. After the interaction the density matrix of the detector is
$\hat{\rho}_{\mathrm{D}}(t)=\tr_{\mathrm{S}}\left\{\hat{U}(t)\left(\hat{\rho}_{
\mathrm{S}}(0)\otimes|\Phi\rangle\langle\Phi|\right)\hat{U}^{\dag}(t)\right\}$
where $\hat{U}(t)$ is the evolution operator. Later, for simplicity we shall
neglect the Hamiltonian of the detector. Then, the evolution operator in the
first-order approximation is \cite{Ruseckas1}
\begin{equation}
\hat{U}(t)\approx\hat{U}_{\mathrm{S}}(t)\left(1+\frac{\lambda\hat{q}}{\mathrm{
i}\hbar}\int_0^t\tilde{A}(t_1)\dd t_1\right)
\end{equation}
where $\hat{U}_{\mathrm{S}}(t)$ is the evolution operator of the unperturbed
system and
$\tilde{A}(t)=\hat{U}_{\mathrm{S}}^{\dag}(t)\hat{A}\hat{U}_{\mathrm{S}}(t)$.
From Eq.~(\ref{eq:sec2:defin}) we obtain that the weak value coincides with the
usual average $\langle\hat{A}\rangle=\tr\{\hat{A}\hat{\rho}_{\mathrm{S}}(0)\}$.

The influence of the weak measurement on the evolution of the measured system
can be made arbitrary small using the small parameter $\lambda$. Therefore, after
the interaction of the measured system with the detector we can try to measure
the second observable $\hat{B}$ using, as usual, the strong measurement. As far as our model
gives the correct result for the value of $A$ averaged over the entire ensemble
of the systems, now we can try to take the average only over the subensemble of the
systems with the given value of the quantity $B$. We measure the momenta
$p_{\mathrm{q}}$ of each measuring device after the interaction with the system.
Subsequently, we perform the final, postselection measurement of $B$ on the
systems of our ensemble. Then we collect the outcomes $p_q$ only of the systems
which have a given value of $B$.

The joint probability that the system has the given value of $B$ \emph{and\/{}}
the detector has the momentum $p_{\mathrm{q}}(t)$ at the time moment $t$ is
$W(B,p_{\mathrm{q}};t)=\tr\left\{|B\rangle\langle B|p_{\mathrm{q}}\rangle\langle
p_{\mathrm{q}}|\hat{\rho}(t)\right\}$, where $|p_{\mathrm{q}}\rangle$ is the
eigenfunction of the momentum operator $\hat{p}_{\mathrm{q}}$. In quantum
mechanics the probability that two quantities simultaneously have definite
values does not always exist. If the joint probability does not exist then the
concept of the conditional probability is meaningless. However, in our case
operators $\hat{B}$ and $|p_{\mathrm{q}}\rangle\langle p_{\mathrm{q}}|$ act in
different spaces and commute, therefore, the probability $W(B,p_{\mathrm{q}};t)$
exists.

Let us define the conditional probability, i.e., the probability that the
momentum of the detector is $p_{\mathrm{q}}$ provided that the system has the
given value of $B$. This probability is given according to Bayes theorem as 
\begin{equation}
W(p_{\mathrm{q}};t|B)=\frac{W(B,p_{\mathrm{q}};t)}{W(B;t)}\label{condprob0}
\end{equation}
 where $W(B;t)=\tr\{|B\rangle\langle B|\hat{\rho}(t)\}$ is the probability that
the system has the given value of $B$. The average momentum of the detector on
condition that the system has the given value of $B$ is 
\begin{equation}
\langle p_{\mathrm{q}}(t)\rangle=\int p_{\mathrm{q}}W(p_{\mathrm{q}};t|B)\dd p_{
\mathrm{q}}\,.\label{condave0}
\end{equation}

>From Eqs.~(\ref{eq:sec2:defin}) and (\ref{condave0}), in the first-order
approximation we obtain the mean value of $A$ on condition that the system
has the given value of $B$ (see for analogy Ref.~\cite{Ruseckas1})
\begin{eqnarray}
\langle A\rangle_B & = &\frac{1}{2\langle B|\hat{\rho}_{\mathrm{
S}}|B\rangle}\left\langle |B\rangle\langle B|\hat{A}+\hat{A}|B\rangle\langle
 B|\right\rangle\nonumber\\
 &  & +\frac{1}{\mathrm{i}\hbar\langle B|\hat{\rho}_{\mathrm{
S}}|B\rangle}\left(\langle q\rangle\langle p_{\mathrm{q}}\rangle-\re\langle\hat{
q}\hat{p}_{\mathrm{q}}\rangle\right)\left\langle\left[|B\rangle\langle B|,\hat{
A}\right]\right\rangle .\label{eq:sec2:x}
\end{eqnarray}
If the commutator $\left[|B\rangle\langle B|,\hat{A}\right]$ in
Eq.~(\ref{eq:sec2:x}) is not zero then, even in the limit of the very weak
measurement, the measured value depends on the particular detector. This fact
means that in such a case we cannot obtain the \emph{definite\/{}} value.
Moreover, the coefficient $(\langle q\rangle\langle
p_{\mathrm{q}}\rangle-\re\langle\hat{q}\hat{p}_{\mathrm{q}}\rangle)$ may be zero
for the specific initial state of the detector, e.g., for the Gaussian
distribution of the coordinate $q$ and momentum $p_{\mathrm{q}}$.

\section{The definition of time under condition that the system is in the given final state}

\label{sec:The-time-on}The most-known problem related to time in quantum mechanics is
the so-called ''tunneling time problem''. We can raise another, more general,
question about the time. Let us consider a system which evolves in time. Let
$\chi$ is one of the observables of the system. During the evolution the value
of $\chi$ changes. We are considering a subset $\Gamma$ of possible values of
$\chi$. The question is \emph{how long the values of $\chi$ belong to this
subset}.

There is another version of the question. If we know the final state of the
system, we may ask how long the values of $\chi$ belong to the subset under
consideration when the system evolves from the initial to a final predefined
state. The question about the tunneling time belongs to such class of the
problems. Really, in the tunneling time problem we ask about the duration the
particle spends in a specified region of the space, and we know that the particle
has tunneled out, i.e., it is on the other side of the barrier. We can expect that
such a question can not always be answered. Here our goal is to obtain the
conditions under which it is possible to answer such question.

One of the possibilities to solve the time problem is to answer what exactly
the word {}``time'' means. The meaning of every physical quantity is determined by the
procedure of its measurement. Therefore, we have to construct a scheme of an
experiment (this can be a \emph{gedanken\/{}} experiment) which measures the
quantity with the properties corresponding to the classical time.

\subsection{The model of the time measurement}

\label{sec:mod}We consider a system evolving with time. Let one of the quantities
describing the system to be $\chi$. Operator $\hat{\chi}$ corresponds to this
quantity. For simplicity we assume that the operator $\hat{\chi}$ has a
continuous spectrum. The case with discrete spectrum will be considered later.

The measuring device interacts with the system only if $\chi$ is in some region
near the point
$\chi_D$, the concrete value of which depends on the detector only. If we want
to measure the time when the
system is in a large region of $\chi$, one has to use many detectors. In the
case of tunneling a similar model was introduced by Steinberg
\cite{steinberg} and developed in our paper \cite{Ruseckas2}. The strong limit
of such a model for analysis of the measurement effect for the quantum jumps has
been used in Ref. \cite{Ruseckas3}.

We shall use the weak measurement concept described in Sec.~\ref{sec:concept}. The
operator $\hat{A}$ will be represented by operator
\begin{equation}
\hat{D}(\chi_{\mathrm{D}})=|\chi_{\mathrm{D}}\rangle\langle\chi_{\mathrm{
D}}|=\delta(\hat{\chi}-\chi_{\mathrm{D}}).\label{delta1}
\end{equation}
It is assumed that after time $t$ the readings of the detectors are collected and averaged.

Hamiltonian (\ref{eq:sec2:11a}) with $\hat{A}$ given by (\ref{delta1})
represents a constant force acting on the detector \textbf{D} when the
quantity $\chi$ is very close to $\chi_{\mathrm{D}}$. This force
induces the change of the detector's momentum. From the classical point of view,
the change of the momentum is proportional to the time the particle spends in
the region around $\chi_{\mathrm{D}}$, and the coefficient of proportionality
equals the force acting on the detector. We assume that the change of the mean
momentum of the detector is proportional to the time the constant force acts on
the detector and that the time the particle spends in the detector's region
coincides with the time the force acts on the detector.

We can replace the $\delta$ function by the narrow rectangle of height $1/L$ and
of width $L$ in the $\chi$ space. From Eq.~(\ref{eq:sec2:11a}) it follows that
the force acting on the detector when the particle is in the region around
$\chi_{\mathrm{D}}$ is $F=-\lambda/L$. The time the particle spends until time
moment $t$ in the unit-length region is 
\begin{equation}
\tau(t)=-\frac{1}{\lambda}\left(\langle p_{\mathrm{q}}(t)\rangle-\langle p_{
\mathrm{q}}\rangle\right)\label{timedef1}
\end{equation}
 where $\langle p_{\mathrm{q}}\rangle$ and $\langle p_{\mathrm{q}}(t)\rangle$
are the mean initial momentum and the momentum after time $t$, respectively. If
one wants to find the period the system spends in the region of the finite width,
one must sum expressions type  (\ref{timedef1}) many times.

When the operator $\hat{\chi}$ has a discrete spectrum, one may ask how long the
quantity $\chi$ has the value $\chi_{\mathrm{D}}$. To answer this question the
detector must interact with the system only when $\chi=\chi_{\mathrm{D}}$. If
this is satisfied, the operator $\hat{D}(\chi_{\mathrm{D}})$ takes the form 
\begin{equation}
\hat{D}(\chi_{\mathrm{D}})=|\chi_{\mathrm{D}}\rangle\langle\chi_{\mathrm{D}}|.
\end{equation}
 The force, acting on the detector in this case equals to $-\lambda$. The
 duration
the quantity $\chi$ has the value $\chi_{\mathrm{D}}$ is given by
Eq.~(\ref{timedef1}), too. Note that now formulae do not depend on spectrum of the
operator $\hat{\chi}$.

\subsection{The dwell time}

\label{sec:mes}To shorten of the notation, the operator
\begin{equation}
\hat{F}(\chi_{\mathrm{D}},t)=\int_0^t\tilde{D}(\chi_{\mathrm{D}},t_1)\dd
 t_1\label{opf1}
\end{equation}
is introduced, where
\begin{equation}
\tilde{D}(\chi_{\mathrm{D}},t)=\hat{U}_{\mathrm{S}}^{\dag}(t)\hat{D}(\chi_{
\mathrm{D}})\hat{U}_{\mathrm{S}}(t).
\end{equation}

After measurement, from the density matrix of the detector, in the first-order
approximation we find that the average change of the detector momentum 
in the time interval $t$ is $-\lambda\langle\hat{F}(\chi_{\mathrm{D}},t)\rangle$.
From Eq.~(\ref{timedef1}) we obtain the dwell time until time moment $t$, 
\begin{equation}
\tau(\chi,t)=\left\langle\hat{F}(\chi,t)\right\rangle .\label{fulltime}
\end{equation}
Then the time spent in the region $\Gamma$ is 
\begin{equation}
\tau(\Gamma;t)=\int_{\Gamma}\tau(\chi,t)\dd\chi=\int_0^t\dd t'\int_{\Gamma}\dd\chi
 P(\chi,t'),\label{eq:dw}
\end{equation}
 where $P(\chi,t')=\langle\tilde{D}(\chi,t)\rangle$ is the probability for the
system to have the value $\chi$ at time moment $t'$.

When $\chi$ is the coordinate of the particle Eq.~(\ref{eq:dw})
yields the well-known expression for the dwell time
\cite{olkhovskyrecami,Ruseckas2}. This time is the average over the entire
ensemble of the systems, regardless of their final states.

A relationship between dwell, transmission, and reflection times recently has been analysed in paper \cite{goto} while in paper \cite{winful} a relation between the group delay and the dwell time for quantum tunneling is derived. It is shown that the group delay is equal to the dwell time plus self-interference delay which depends on the dispersion outside the barrier. The analysis shows that there is nothing superluminal in quantum tunneling and the Hartman effect for tunneling quantum particles can be explained by the saturation of the integrated probability density under the barrier.

\subsection{The definition of time under condition that the system is in the given final state}

\label{sec:condprob}Further, the case when the final state of
the system is known will be considered . We may ask \emph{how long the values of $\chi$ belong to
the subset under consideration, $\Gamma$, on condition that the system evolves
to the definite final state $f$}. More specifically, we might know that the final
state of the system belongs to a certain subspace $\mathcal{H}_{\mathrm{f}}$
of system's Hilbert space.
The projection operator that projects the vectors from the Hilbert space of the
system into the subspace $\mathcal{H}_{\mathrm{f}}$ of the final states will be
denoted
$\hat{P}_{\mathrm{f}}$. As far as the considered model gives the correct result for the
time averaged over the entire ensemble of the systems, we can try to take the
average only over the subensemble of the systems with the given final states. At
first, the momenta $p_{\mathrm{q}}$ of each measuring device after the
interaction with the system are measured. Subsequently, we perform the final, postselection
measurement on the systems of the ensemble. Then we collect the outcomes $p_q$
only of those systems the final state of which turns out to belong to the subspace
$\mathcal{H}_{\mathrm{f}}$.

Using Eq.~(\ref{eq:sec2:x}) in Sec.~\ref{sec:concept} we obtain the duration,
on condition that the final state of the system belongs to the subspace
$\mathcal{H}_{\mathrm{f}}$ \cite{Ruseckas1},
\begin{eqnarray}
\tau_{\mathrm{f}}(\chi,t) & = &\frac{1}{2\langle\tilde{P}_{\mathrm{
f}}(t)\rangle}\left\langle\tilde{P}_{\mathrm{f}}(t)\hat{F}(\chi,t)+\hat{
F}(\chi,t)\tilde{P}_{\mathrm{f}}(t)\right\rangle\nonumber\\
 &  & +\frac{1}{\mathrm{i}\hbar\langle\tilde{P}_{\mathrm{
f}}(t)\rangle}\left(\langle q\rangle\langle p_{\mathrm{q}}\rangle
-\re\langle\hat{q}\hat{p}_{\mathrm{q}}\rangle\right)\left\langle\left[\tilde{
P}_{\mathrm{f}}(t),\hat{F}(\chi,t)\right]\right\rangle .\label{condtime}
\end{eqnarray}

Eq.~(\ref{condtime}) consists of two terms, thus, we can introduce two expressions
with the dimension of time 
\begin{eqnarray}
\tau_{\mathrm{f}}^{(1)}(\chi,t) & = &\frac{1}{2\langle\tilde{P}_{\mathrm{
f}}(t)\rangle}\left\langle\tilde{P}_{\mathrm{f}}(t)\hat{F}(\chi,t)+\hat{
F}(\chi,t)\tilde{P}_{\mathrm{f}}(t)\right\rangle ,\label{timere}\\
\tau_{\mathrm{f}}^{(2)}(\chi,t) & = &\frac{1}{2\mathrm{i}\langle\tilde{P}_{
\mathrm{f}}(t)\rangle}\left\langle\left[\tilde{P}_{\mathrm{f}}(t),\hat{
F}(\chi,t)\right]\right\rangle .\label{timeim}
\end{eqnarray}
Then, the time the system spends in the subset $\Gamma$ on condition that the
final state of the system belongs to the subspace $\mathcal{H}_{\mathrm{f}}$ can
be rewritten in the form
\begin{equation}
\tau_{\mathrm{f}}(\chi,t)=\tau_{\mathrm{f}}^{(1)}(\chi,t)+\frac{2}{
\hbar}\left(\langle q\rangle\langle p_{\mathrm{q}}\rangle-\re\langle\hat{q}\hat{
p}_{\mathrm{q}}\rangle\right)\tau_{\mathrm{f}}^{(2)}(\chi,t).\label{condtime2}
\end{equation}
 The quantities $\tau_{\mathrm{f}}^{(1)}(\chi,t)$ and
$\tau_{\mathrm{f}}^{(2)}(\chi,t)$ are related to the real and imaginary parts of
the complex time introduced by Sokolovski \textit{et al\/{}}
\cite{sokolovskibaskin}. In our model the quantity $\tau_{\mathrm{f}}(\chi,t)$
is real, contrary to the complex-time approach. The components of time
$\tau_{\mathrm{f}}^{(1)}$ and $\tau_{\mathrm{f}}^{(2)}$ are real, too.
Therefore, this time can be interpreted as the duration of an event.

If the commutator $\left[\tilde{P}_{\mathrm{f}}(t),\hat{F}(\chi,t)\right]$ in
Eq.~(\ref{condtime}) is not zero then, even in the limit of very weak
measurement, the measured value depends on the particular detector used. This 
means that in such a case we cannot obtain a \emph{definite\/{}} value for the
conditional time. Moreover, the coefficient $(\langle q\rangle\langle
p_{\mathrm{q}}\rangle-\re\langle\hat{q}\hat{p}_{\mathrm{q}}\rangle)$ may be zero
for the specific initial state of the detector, e.g., for the Gaussian
distribution of the coordinate $q$ and momentum $p_{\mathrm{q}}$.

The conditions to determine the time uniquely in a case when
the final state of the system is known takes, thus, the form
\begin{equation}
\left[\tilde{P}_{\mathrm{f}}(t),\hat{F}(\chi,t)\right]=0\label{eq:poscond}
\end{equation}
which can be understood from on general principles of the quantum
mechanics, too. Now, we ask \emph{how long the values of $\chi$ belong to a
certain subset when the system evolves to the given final state} under
assumption that the final state of the system is known with
certainty. In addition, we want to have some
information about the values of the quantity $\chi$. However, if the
final state is known with certainty, we may not know the values of $\chi$ in the past
and, vice versa, if we know something about $\chi$, we may not definitely
determine the final state. Therefore, in such a case the question about the time
when the system evolves to the given final state cannot be answered definitely
and the conditional time has no reasonable meaning.

The quantity $\tau_{\mathrm{f}}(t)$ according to Eqs.~(\ref{condtime}) and
(\ref{timere}) has many properties of the classical time. So, if the final
states $\{f\}$ constitute the full set, then the corresponding projection
operators obey the equality of completeness $\sum_f\hat{P}_{\mathrm{f}}=1$.
Then,
from Eq.~(\ref{condtime}) we obtain the expression 
\begin{equation}
\sum_f\langle\tilde{P}_{\mathrm{f}}(t)\rangle\tau_{\mathrm{
f}}(\chi,t)=\tau(\chi,t).\label{clasprop}
\end{equation}
 The quantity $\langle\tilde{P}_{\mathrm{f}}(t)\rangle$ is the probability that
the system at the time $t$ is in the state $f$. Eq.~(\ref{clasprop}) shows that
the full duration equals the average over all possible final states, as it is
a case in the classical physics. From Eq.~(\ref{clasprop}) and
Eqs.~(\ref{timere}), (\ref{timeim}) it follows 
\begin{eqnarray}
\sum_f\langle\tilde{P}_{\mathrm{f}}(t)\rangle\tau_{\mathrm{f}}^{(1)}(\chi,t) & =
 &\tau(\chi,t),\\
\sum_f\langle\tilde{P}_{\mathrm{f}}(t)\rangle\tau_{\mathrm{f}}^{(2)}(\chi,t) & =
 & 0.
\end{eqnarray}
 We suppose that the quantities $\tau_{\mathrm{f}}^{(1)}(\chi,t)$ and
$\tau_{\mathrm{f}}^{(2)}(\chi,t)$ can be useful even in the case when the time
has no definite value, since in the tunneling time problem the quantities
(\ref{timere}) and (\ref{timeim}) correspond to real and imaginary parts of
the complex time, respectively \cite{Ruseckas2}.

The eigenfunctions of the operator $\hat{\chi}$ constitute the full set
$\int|\chi\rangle\langle\chi|\dd\chi=1$, where the integral must be replaced by
the sum for the discrete spectrum of the operator $\hat{\chi}$. From
Eqs.~(\ref{delta1}), (\ref{opf1}), (\ref{condtime}) we obtain the equality 
\begin{equation}
\int\tau_{\mathrm{f}}(\chi,t)\dd\chi=t,\label{eq:eqx}
\end{equation}
which shows that the time during which the quantity $\chi$ has any
value equals to $t$, as it is in the classical physics.

\subsection{Example: two-level system}

\label{sec:exampl}The obtained formalism can be applied to the tunneling time
problem \cite{Ruseckas2}. In this section, however, we will consider a simpler
system than the tunneling particle, i.e., a two-level system. The system is
forced by the perturbation $V$ that causes the jumps from one state to another.
The time the system is in a given state will be calculated.

The Hamiltonian of this system is 
\begin{equation}
\hat{H}=\hat{H}_0+\hat{V}
\end{equation}
 where $\hat{H}_0=\hbar\omega\hat{\sigma}_3/2$ is the Hamiltonian of the
unperturbed system and $\hat{V}=v\hat{\sigma}_{+}+v^*\hat{\sigma}_{-}$ is the
perturbation. Here $\hat{\sigma}_1,\hat{\sigma}_2,\hat{\sigma}_3$ are Pauli matrices and
$\sigma_{\pm}=\frac{1}{2}(\hat{\sigma}_1\pm i\hat{\sigma}_2)$. The Hamiltonian $\hat{H}_0$
has two eigenfunctions $|0\rangle$ and $|1\rangle$ with the eigenvalues
$-\hbar\omega/2$ and $\hbar\omega/2$, respectively. The initial state of the
system assumed to be $|0\rangle$.

>From Eq.~(\ref{fulltime}) we obtain the times the system spends in the energy
levels $0$ and $1$, respectively, 
\begin{eqnarray}
\tau(0,t) & = &\frac{1}{2}\left(1+\frac{\omega^2}{\Omega^2}\right)t+\frac{1}{
2\Omega}\sin(\Omega t)\left(1-\frac{\omega^2}{\Omega^2}\right),\label{eq:31}\\
\tau(1,t) & = &\frac{1}{2}\left(1-\frac{\omega^2}{\Omega^2}\right)t-\frac{1}{
2\Omega}\sin(\Omega t)\left(1-\frac{\omega^2}{\Omega^2}\right)\label{eq:32}
\end{eqnarray}
 where $\Omega=\sqrt{\omega^2+4(|v|/\hbar)^2}$. From Eqs. (\ref{timere}) and
(\ref{timeim}) we can obtain the conditional time. The components (\ref{timere})
and (\ref{timeim}) of the time the system spends in
the level $0$ under condition that the final state after measurement is $|1\rangle$ are 
\begin{eqnarray}
\tau_1^{(1)}(0,t) & = &\frac{t}{2},\label{eq:34}\\
\tau_1^{(2)}(0,t) & = &\frac{\omega}{2\Omega}\left(1-t\cot\left(\frac{\Omega}{
2}t\right)\right).
\end{eqnarray}
 When $\Omega t=2\pi n$, where $n\in\mathbb{Z}$, the quantity $\tau_1^{(2)}(0,t)$
tends to infinity. This happens because at these time moments the system is in
the state
$|1\rangle$ with the probability $0$, and one cannot consider the interaction with
the detector as very weak.

\begin{figure}
\begin{center}
\includegraphics[width=0.60\textwidth]{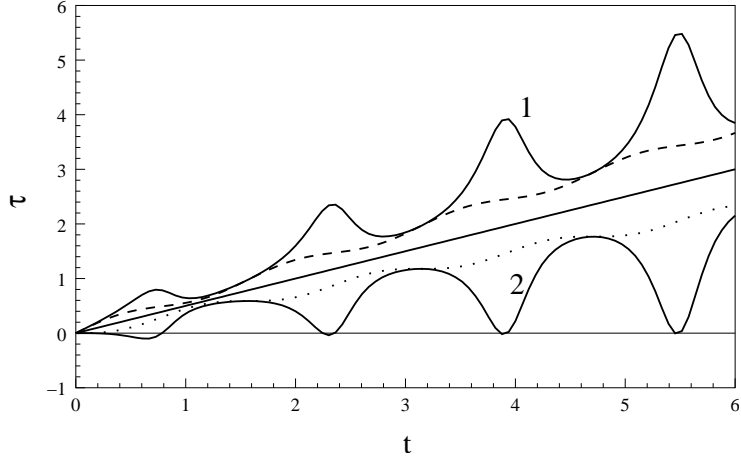}
\end{center}
\caption{The times the system spends in the energy level $0$, $\tau(0,t)$
(dashed line), and level $1$, $\tau(1,t)$ (dotted line), according to
Eqs.~(\ref{eq:31}) and (\ref{eq:32}), respectively. The quantity
$\tau_1^{(1)}(0,t)$, Eq.~(\ref{eq:34}), is shown as solid straight line. The
quantities $\tau_0^{(1)}(0,t)$ and $\tau_0^{(1)}(1,t)$, shown by curves 1 and 2,
were calculated
according to Eqs.~(\ref{eq:28}) and (\ref{eq:30}), respectively. The parameters
are $\omega=2$, $\Omega=4$.}
\label{dwt}
\end{figure}
On the other hand, the components of the time the system spends in level $0$
under condition that the final state
is $|0\rangle$ are 
\begin{eqnarray}
\tau_0^{(1)}(0,t) & = &\frac{\left(1+3\frac{\omega^2}{\Omega^2}\right)t+\left(1
-\frac{\omega^2}{\Omega^2}\right)\left(\frac{2}{\Omega}\sin(\Omega t)
+t\cos(\Omega t)\right)}{2\left(\left(1+\frac{\omega^2}{\Omega^2}\right)+\left(1
-\frac{\omega^2}{\Omega^2}\right)\cos(\Omega t)\right)},\label{eq:28}\\
\tau_0^{(2)}(0,t) & = &\frac{\frac{\omega}{\Omega}\left(1-\frac{\omega^2}{
\Omega^2}\right)\sin\left(\frac{\Omega}{2}t\right)\left(t\cos\left(\frac{
\Omega}{2}t\right)-\frac{2}{\Omega}\sin\left(\frac{\Omega}{2}t\right)\right)}{
2\left(\left(1+\frac{\omega^2}{\Omega^2}\right)+\left(1-\frac{\omega^2}{
\Omega^2}\right)\cos(\Omega t)\right)}.\label{eq:29}
\end{eqnarray}

\begin{figure}
\begin{center}\includegraphics[width=0.60\textwidth]{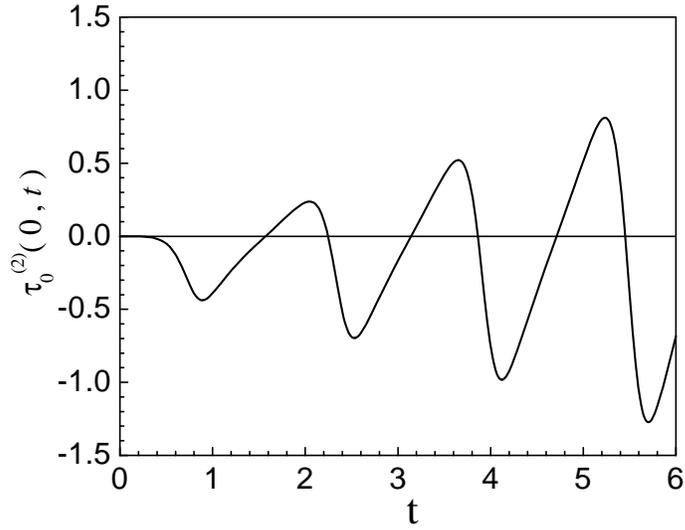}
\end{center}
\caption{The quantity $\tau_0^{(2)}(0,t)$, Eq.~(\ref{eq:29}). The parameters are
the same as in Fig.~\ref{dwt}.}
\label{ti00}
\end{figure}
The time the system spends in level $1$ under condition
that the final state is $|0\rangle$ may be expressed as 
\begin{equation}
\tau_0^{(1)}(1,t)=\frac{\left(1-\frac{\omega^2}{\Omega^2}\right)\left(t
+t\cos(\Omega t)-\frac{2}{\Omega}\sin(\Omega t)\right)}{2\left(\left(1+\frac{
\omega^2}{\Omega^2}\right)+\left(1-\frac{\omega^2}{\Omega^2}\right)\cos(\Omega
 t)\right)}.\label{eq:30}
\end{equation}

The quantities $\tau(0,t)$, $\tau(1,t)$, $\tau_1^{(1)}(0,t)$,
$\tau_0^{(1)}(0,t)$ and $\tau_0^{(1)}(1,t)$ are shown in Fig.~\ref{dwt}. The
quantity $\tau_0^{(2)}(0,t)$ is shown in Fig.~\ref{ti00}. Note that the partial
durations
at the given final state are not necessarily monotonic as it is with the full
duration, because the final state at different time moments can be reached by
different paths. We can interpret the quantity $\tau_0^{(1)}(0,t)$ as the time
the system spends in the level $0$ on condition that the final state is
$|0\rangle$, but at certain time moments this quantity is greater than $t$. In
such cases the
quantity $\tau_0^{(1)}(1,t)$ becomes negative at certain times. This is
a consequence of the fact that for the system under consideration the
condition (\ref{eq:poscond}) is not fulfilled. The peculiarities of the behavior
of the conditional times show that it is impossible to decompose the
unconditional time into two components having all classical properties of the
time.

\section{Tunneling time}

\label{sec:Tunneling-time}The best-known problem of time in quantum mechanics is
the so-called ''tunneling time problem''. This problem is still the subject of
much controversy, since numerous theories contradict each other in their
predictions for {}``the tunneling time''. Many of the theoretical approaches can
be divided into three categories. First, one can study the evolution of the wave
packets through the barrier and get the phase time. However, the correctness of
the definition of this time is highly questionable \cite{buttikerlandauer}.
Another approach is based on the determination of a set of dynamic paths, i.e.,
the calculation of the time the different paths spend in the barrier and
averaging over the set of the paths. The paths can be found from the Feynman
path integral formalism \cite{sokolovskibaskin}, from the Bohm approach
\cite{leavens1,leavens2,leavens3,leavens4,leavens5,aharonoverez}, or from
the Wigner distribution \cite{muga}. The third class uses a physical clock which is used for
determination of the time elapsed during the tunneling (B\"{u}ttiker and
Landauer used an oscillatory barrier \cite{buttikerlandauer}, Baz' suggested the
Larmor time \cite{baz}). One more approach is based on a model for tunneling
based on stochastic interpretation of quantum mechanics \cite{42a,42b,42c,42d}.

The problems rise also from the fact that the arrival time of a particle to a
definite spatial point is a classical concept. Its quantum counterpart is
problematic even for the free particle case. In classical mechanics, for the
determination of the time the particle spends moving along a certain trajectory,
one has to measure the position of the particle at two different moments of
time. In quantum mechanics this procedure does not work. From Heisenberg's
uncertainty principle it follows that we cannot measure the position of a
particle without alteration of its momentum. To determine exactly the arrival
time of a particle, one has to measure the position of the particle with great
precision. Because of the measurement, the momentum of the particle will have a
big uncertainty and the second measurement will be indefinite. If we want to ask
about the time in quantum mechanics, we need to define the procedure of
measurement. We can measure the position of the particle only with a finite
precision and get a distribution of the possible positions. Applying such a
measurement, we can expect to obtain not a single value of the traversal time
but a distribution of times.

In paper \cite{42e} the tunneling time distribution for photon tunneling is
analysed theoretically as a space-time correlation phenomenon between the
emission and absorption of a photon on the two sides of a barrier. The analysis
is based on an appropriate counting rate formula derived at first order in the
photon-detector interaction and used in treating space-time correlations between
photons.

There are two different but related questions
connected with the tunneling time problem \cite{dumontmarchioro}:
\begin{enumerate}
\item[(i)] How much time does the tunneling particle spend
under the barrier?
\item[(ii)] At what time does the particle arrive at the point
behind the barrier?
\end{enumerate}
There have been many attempts to answer
these questions. However, there are several papers showing that according to
quantum mechanics the question (i) makes no sense
\cite{dumontmarchioro,sokolovskiconnor,yamada,BSM}. Our goal is to investigate
the possibility to determine the tunneling time using weak measurements.

\subsection{Determination of the tunneling time}

\label{secimpos}To answer the question of \textit{how much time does the
tunneling particle spends under the barrier,} we need a criterion of the
tunneling. The following criterion is accepted: the particle had tunneled in the
case when it was in front of the barrier at first and later it was found behind
the barrier. We shall require that the mean energy of the particle and the energy
uncertainty should be less than the height of the barrier. Following this
criterion, the operator corresponding to the {}``tunneling-flag''
observable is introduced
\begin{equation}
\hat{f}_T(X)=\Theta(\hat{x}-X),\label{tunflag}
\end{equation}
where $\Theta$ is the Heaviside unit step function and $X$ is a point behind
the barrier. This operator projects the wave function onto the subspace of
functions localized behind the barrier. The operator has two eigenvalues: $0$
and $1$. The eigenvalue $0$ corresponds to the fact that the particle has not
tunneled out, while the eigenvalue $1$ corresponds to the appearance of particle
behind the barrier.

We will work with the Heisenberg representation. In this representation, the
tunneling flag operator becomes 
\begin{equation}
\tilde{f}_T(t,X)=\exp\left(\frac{i}{\hbar}\hat{H}t\right)\hat{f}_T(X)\exp\left(
-\frac{i}{\hbar}\hat{H}t\right).\label{tunflagheis}
\end{equation}
To take into account all the tunneled particles, the limit
$t\rightarrow+\infty$ must be taken. So, the {}``tunneling-flag'' observable in
the Heisenberg picture is represented by the operator
$\tilde{f}_T(\infty,X)=\lim_{t\rightarrow+\infty}\tilde{f}_T(t,X)$. One can
obtain the explicit expression for this operator.

The operator $\tilde{f}_T(t,X)$ obeys the standard equation 
\begin{equation}
i\hbar\frac{\partial}{\partial t}\tilde{f}_T(t,X)=\left[\tilde{f}_T(t,X),\hat{
H}\right].\label{eq:ft}
\end{equation}
 The commutator in Eq.~(\ref{eq:ft}) may be expressed as 
\[
\left[\tilde{f}_T(t,X),\hat{H}\right]=\exp\left(\frac{i}{\hbar}\hat{
H}t\right)\left[\hat{f}_T(X),\hat{H}\right]\exp\left(-\frac{i}{\hbar}\hat{
H}t\right).
\]
 If the Hamiltonian has the form $\hat{H}=\frac{1}{2M}\hat{p}^2+V(\hat{x})$,
then the commutator becomes
\begin{equation}
\left[\hat{f}_T(X),\hat{H}\right]=i\hbar\hat{J}(X),
\end{equation}
where $\hat{J}(X)$ is the probability flux operator,
\begin{equation}
\hat{J}(x)=\frac{1}{2M}\left(|x\rangle\langle x|\hat{p}+\hat{p}|x\rangle\langle
 x|\right).\label{probflux}
\end{equation}
Therefore, the following equation for the commutator can be written 
\begin{equation}
\left[\tilde{f}_T(t,X),\hat{H}\right]=i\hbar\tilde{J}(X,t).\label{eq:com}
\end{equation}
The initial condition for the function $\tilde{f}(t,X)$ may be defined as 
\[
\tilde{f}_T(t=0,X)=\hat{f}_T(X).
\]
 From Eqs.~(\ref{eq:ft}) and~(\ref{eq:com}) we obtain the equation for the
evolution of the tunneling-flag operator 
\begin{equation}
i\hbar\frac{\partial}{\partial t}\tilde{f}_T(t,X)=i\hbar\tilde{J}(X,t).
\label{eq:evol}
\end{equation}
 From Eq.~(\ref{eq:evol}) and the initial condition, an explicit expression for
the tunneling-flag operator follows
\begin{equation}
\tilde{f}_T(t,X)=\hat{f}_T(X)+\int_0^t\tilde{J}(X,t_1)\dd t_1.
\label{tunflagflux}
\end{equation}

In the already mentioned question of \emph{how much time does the tunneling particle spend under
the barrier}, we shall be interested in those particles, which we know with certainty have
tunneled out. In addition, we want to have some information about the location of
the particle. However, one may ask whether the quantum mechanics allows one to have the information
about the tunneling and location simultaneously? The projection operator 
\begin{equation}
\hat{D}(\Gamma)=\int\limits_{\Gamma}|x\rangle\langle x|\dd x\label{posop}
\end{equation}
 represents the probability for the particle to be in the region $\Gamma$. Here
$|x\rangle$ is the eigenfunction of the coordinate operator. In the Heisenberg
representation this operator takes the form
\begin{equation}
\tilde{D}(\Gamma,t)=\exp\left(\frac{i}{\hbar}\hat{H}t\right)\hat{
D}(\Gamma)\exp\left(-\frac{i}{\hbar}\hat{H}t\right).\label{posopheis}
\end{equation}
From Eqs.~(\ref{probflux}),~(\ref{tunflagflux}), and~(\ref{posopheis}) we see
that the operators $\tilde{D}(\Gamma,t)$ and $\tilde{f}_T(\infty,X)$ in general
do not commute. This means that we cannot simultaneously have the information
about the tunneling and location of the particle. If we know with certainty that
the particle has tunneled out then we can say nothing about its location in the
past,
and if we know something about the location of the particle, we cannot determine
definitely whether the particle has tunnel out. Therefore, the question of
\textit{how much time does the tunneling particle spend under the barrier}
cannot have definite answer, if the question is so posed that its precise
definition requires the existence of the joint probability that the particle is
found in $\Gamma$ at time $t$ and whether or not it is found on the right side
of the barrier at a sufficiently later time. A similar analysis has been
performed in Ref. \cite{BSM}. It has been shown that, due to noncommutability of
operators, there exist no unique decomposition of the dwell time.

This conclusion is, however, not negative altogether. We know that
$\int_{-\infty}^{+\infty}|x\rangle\langle x|\dd x=1$ and
$\left[1,\tilde{f}_T(\infty,X)\right]=0$. Therefore, if the region $\Gamma$ is
large enough, one has a possibility to answer the question about the tunneling
time.

>From the fact that the operators $\tilde{D}(\Gamma,t)$ and
$\tilde{f}_T(\infty,X)$ do not commute we can predict that the measurement of
the tunneling time will yield a value dependent on the particular detection
scheme. We shall assume the detector is made so that it yields some value. But if we try to
measure noncommuting observables, the measured values depend on the interaction
between the detector and the measured system. So, in the definition of the
Larmor time there is a dependence on the type of boundary attributed to the
magnetic-field region \cite{olkhovskyrecami}.

\subsection{The model of the time measurement}

\label{secmod}We consider a model for the tunneling time measurement which is
somewhat similar to the gedanken experiment used to obtain the Larmor
time but is simpler and more transparent. This model was proposed by
Steinberg \cite{steinberg}, however, it was treated in a nonstandard way,
introducing complex probabilities. Here we shall use only the formalism of the
standard quantum mechanics.

\begin{figure}
\begin{center}\includegraphics[width=0.60\textwidth]{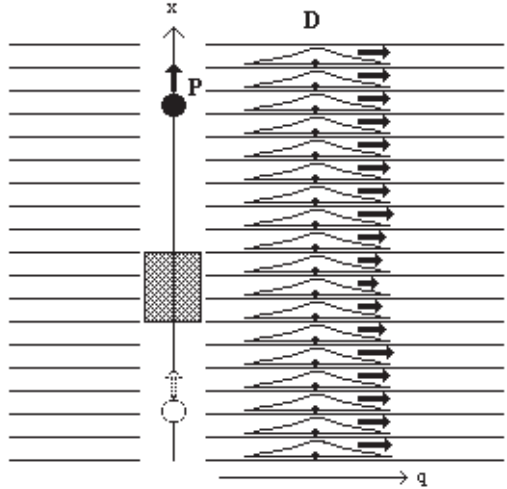}
\end{center}
\caption{The configuration of the measurement of the tunneling time. The
particle \textbf{P} is tunneling along $x$ coordinate and weakly interacts with
detectors \textbf{D}. The barrier is represented by the hatched rectangle. The
interaction with the individual detectors occurs only in the narrow region limited
by the horizontal lines. The changes of the momenta of the detectors are
represented by arrows.}
\label{config}
\end{figure}

Our system consists of a particle \textbf{P} and a number of detectors \textbf{D}
\cite{Ruseckas2}. Each detector interacts with the particle only in the narrow
region of space. The configuration of the system is shown in Fig.~\ref{config}.
When the interaction of the particle with the detectors is weak, the detectors
do not influence the state of the particle. Therefore, we can analyze the action
of the detectors separately. This model is a particular case of time measurement
presented in Sec.~\ref{sec:mod}, with $\chi_D$ being the position of the
detector $x_D$. Similar calculations were done for detector's position rather
than momentum by Iannaccone \cite{iann}.

In the moment $t=0$ the particle is before the barrier, therefore, $\langle
x|\rho_P(0)|x^{\prime}\rangle\neq0$ only when $x<0$ and $x^{\prime}<0$, where
$\hat{\rho}_P(0)$ is the density matrix of the particle \textbf{P}.

\subsection{Measurement of the dwell time}

\label{secdwell}As in Sec.~\ref{sec:mes} we obtain the time the particle spends
in the unit length region between time instances $0$ and $t$
\begin{equation}
\tau^{\mathrm{Dw}}(x,t)=\langle\hat{F}(x,t)\rangle.\label{dwelltime}
\end{equation}
The time spent in the space region restricted by the coordinates $x_1$ and $x_2$
is 
\begin{equation}
t^{\mathrm{Dw}}(x_2,x_1)=\int_{x_1}^{x_2}\tau^{\mathrm{
Dw}}(x,t\rightarrow\infty)\dd x=\int_{x_1}^{x_2}\dd x\int_0^{\infty}\rho(x,t)\dd
 t
\end{equation}
which is a well-known expression for the dwell time \cite{olkhovskyrecami}: the
dwell time is the average over an entire ensemble of particles regardless
whether they tunneled or not. The expression for the dwell time obtained in our model is the
same as the well-known expression obtained by other authors. Therefore, we can
expect that our model can yield a reasonable expression for the tunneling time
as well.

\subsection{Conditional probabilities and the tunneling time}

\label{seccondprob}Having seen that our model is capable to give the time averaged over
entire ensemble of the particles, let us now take the average over the
subensemble of the tunneled particles only. This will be done similarly to
Sec.~\ref{sec:condprob} with $\hat{P}_{\mathrm{f}}$ replaced by the tunneling-flag
operator $\hat{f}_T(X)$ defined by Eq.~(\ref{tunflag}). From
Eq.~(\ref{condtime}) we obtain the duration the tunneled particle spends in the unit
length region around $x$ until time $t$ \cite{Ruseckas2}
\begin{eqnarray}
\tau(x,t) & = &\frac{1}{2\langle\tilde{f}_T(t,X)\rangle}\left\langle\tilde{
f}_T(t,X)\hat{F}(x,t)+\hat{F}(x,t)\tilde{f}_T(t,X)\right\rangle\nonumber\\
 &  & +\frac{1}{i\hbar\langle\tilde{f}_T(t,X)\rangle}\left(\langle
 q\rangle\langle p_q\rangle-\re\langle\hat{q}\hat{
p}_q\rangle\right)\left\langle\left[\tilde{f}_T(t,X),\hat{
F}(x,t)\right]\right\rangle .\label{tuntime}
\end{eqnarray}
 The obtained expression~(\ref{tuntime}) for the tunneling time is real,
contrary to the complex-time approach. It should be noted that this expression,
even in the limit of weak measurement, depends on a particular
detector. If the commutator $[\tilde{f}_T(t,X),\hat{F}(x,t)]$ is zero, the time
has a well-defined value. If the commutator is not zero, only the integral of this
expression over a large region has meaning of an asymptotic time related to
the large region as we will see in Sec. \ref{secasympt}.

Equation (\ref{tuntime}) can be rewritten as a sum of two terms, the first term
being independent and the second dependent on the detector, i.e., 
\begin{equation}
\tau(x,t)=\tau^{\mathrm{Tun}}(x,t)+\frac{2}{\hbar}\left(\langle q\rangle\langle
 p_q\rangle-\re\langle\hat{q}\hat{p}_q\rangle\right)\tau_{\mathrm{corr}}^{
\mathrm{Tun}}(x,t)
\end{equation}
 where
\begin{eqnarray}
\tau^{\mathrm{Tun}}(x,t) & = &\frac{1}{2\langle\tilde{
f}_T(t,X)\rangle}\left\langle\tilde{f}_T(t,X)\hat{F}(x,t)+\hat{F}(x,t)\tilde{
f}_T(t,X)\right\rangle ,\label{tunteimre}\\
\tau_{\mathrm{corr}}^{\mathrm{Tun}}(x,t) & = &\frac{1}{2\, i\langle\tilde{
f}_T(t,X)\rangle}\left\langle\left[\tilde{f}_T(t,X),\hat{
F}(x,t)\right]\right\rangle .\label{tuntimeim}
\end{eqnarray}
The quantities $\tau^{\mathrm{Tun}}(x,t)$ and
$\tau_{\mathrm{corr}}^{\mathrm{Tun}}(x,t)$ are independent of the detector.

In order to separate the tunneled and reflected particles the
limit $t\rightarrow\infty$ should be taken. Otherwise, the particles that
tunneled after the time
$t$ will not contribute. If we introduce the operators
\begin{eqnarray}
\hat{F}(x) & = &\int_0^{\infty}\tilde{D}(x,t_1)\dd t_1,\label{opefinf}\\
\hat{N}(x) & = &\int_0^{\infty}\tilde{J}(x,t_1)\dd t_1.\label{opN}
\end{eqnarray}
then from Eq.~(\ref{tunflagflux}) follows that the operator
$\tilde{f}_T(\infty,X)$ is $\hat{f}_T(X)+\hat{N}(X)$. If the
particle  before the barrier is initially, then 
\[
\hat{f}_T(X)\hat{\rho}_P(0)=\hat{\rho}_P(0)\hat{f}_T(X)=0.
\]
In the limit $t\rightarrow\infty$ tunneling times become
\begin{eqnarray}
\tau^{\mathrm{Tun}}(x) & = &\frac{1}{2\langle\hat{N}(X)\rangle}\left\langle\hat{
N}(X)\hat{F}(x)+\hat{F}(x)\hat{N}(X)\right\rangle ,\label{tuntimereinf}\\
\tau_{\mathrm{corr}}^{\mathrm{Tun}}(x) & = &\frac{1}{2\, i\langle\hat{
N}(X)\rangle}\left\langle\left[\hat{N}(X),\hat{F}(x)\right]\right\rangle
 .\label{tuntimeiminf}
\end{eqnarray}

Let us define an {}``asymptotic time'' as the integral of $\tau(x,\infty)$ over
a wide region containing the barrier. Since the integral of
$\tau_{\mathrm{corr}}^{\mathrm{Tun}}(x)$ is very small compared to that of
$\tau^{\mathrm{Tun}}(x)$ as we shall see later, the asymptotic time is
effectively the integral of $\tau^{\mathrm{Tun}}(x)$ only. This allows us to
identify $\tau^{\mathrm{Tun}}(x)$ as {}``the density of the tunneling time''.

In many cases for the simplification of mathematics it is common to write the
integrals over time as the integrals from $-\infty$ to $+\infty$. In our model
we cannot, without additional assumptions, integrate Eqs.~(\ref{opefinf}),
(\ref{opN}) from $-\infty$ because the negative time means the motion of the
particle to the initial position. If some particle in the initial wave packet
had negative momentum then in the limit $t\rightarrow-\infty$ it was behind the
barrier and contributed to the tunneling time.

\subsection{Properties of the tunneling time}

\label{secprop}As stated, the question of \textit{how much time does a tunneling
particle spend under the barrier} has no exact answer. We
can determine only the time the tunneling particle spends in a large region
containing the barrier. In our model this time is expressed as an integral of
quantity~(\ref{tuntimereinf}) over this region. In order to determine the
properties of this integral it is useful to determine the properties of the
integrand.

To be able to expand the range of integration over time to $-\infty$, it is
necessary to have the initial wave packet far to the left from the points under
the investigation and this wave packet must consist only of the waves moving in
the positive direction.

It is convenient to perform calculations in the energy representation.
Eigenfunctions of the Hamiltonian $\hat{H}_P$ are $|E,\alpha\rangle$, where
$\alpha=\pm1$. The sign '$+$' or '$-$' corresponds to the positive or negative
initial direction of the wave, respectively. Outside the barrier these
eigenfunctions are \label{eigenfunct}
\begin{eqnarray}
\langle x|E,+\rangle & = &\left\{
\begin{array}{l}
\sqrt{\frac{M}{2\pi\hbar p_E}}\left(\exp\left(\frac{i}{\hbar}p_Ex\right)
+r(E)\exp\left(-\frac{i}{\hbar}p_Ex\right)\right),\quad x<0,\\
\sqrt{\frac{M}{2\pi\hbar p_E}}t(E)\exp\left(\frac{i}{\hbar}p_Ex\right),\quad
 x>L,
\end{array}\right.\label{eigenplus}\\
\langle x|E,-\rangle & = &\left\{
\begin{array}{l}
\sqrt{\frac{M}{2\pi\hbar p_E}}t(E)\exp\left(-\frac{i}{\hbar}p_Ex\right),\quad
 x<0,\\
\sqrt{\frac{M}{2\pi\hbar p_E}}\left(\exp\left(-\frac{i}{\hbar}p_Ex\right)-\frac{
t(E)}{t^*(E)}r^*(E)\exp\left(\frac{i}{\hbar}p_Ex\right)\right),\quad x>L
\end{array}
\right.
\label{eigenminus}
\end{eqnarray}
 where $t(E)$ and $r(E)$ are transmission and reflection amplitudes
respectively, and
\begin{equation}
p_E=\sqrt{2ME}.\label{impuls}
\end{equation}
$M$ is the mass of the particle. The barrier is in the region between
$x=0$ and $x=L$. These eigenfunctions are orthonormal, i.e.,
\begin{equation}
\langle E,\alpha|E^{\prime},\alpha^{\prime}\rangle=\delta_{\alpha,\alpha^{
\prime}}\delta(E-E^{\prime}).\label{orthogonality}
\end{equation}
 The evolution operator is 
\[
\hat{U}_P(t)=\sum_{\alpha}\int_0^{\infty}|E,\alpha\rangle\langle
 E,\alpha|\exp\left(-\frac{i}{\hbar}Et\right)\dd E.
\]
 Then the operator $\hat{F}(x)$ assumes the form 
\[
\hat{F}(x)=\int_{-\infty}^{\infty}dt_1\sum_{\alpha,\alpha^{
\prime}}\int\!\!\!\int\dd E\,\dd E^{\prime}|E,\alpha\rangle\langle
 E,\alpha|x\rangle\langle x|E^{\prime},\alpha^{\prime}\rangle\langle E^{
\prime},\alpha^{\prime}|\exp\left(\frac{i}{\hbar}\left(E-E^{
\prime}\right)t_1\right)
\]
 where the integral over the time yields $2\pi\hbar\delta(E-E^{\prime})$ and,
therefore, 
\[
\hat{F}(x)=2\pi\hbar\sum_{\alpha,\alpha^{\prime}}\int\dd
 E|E,\alpha\rangle\langle E,\alpha|x\rangle\langle x|E,\alpha^{
\prime}\rangle\langle E,\alpha^{\prime}|.
\]
Similarly, we find
\[
\hat{N}(x)=2\pi\hbar\sum_{\alpha,\alpha^{\prime}}\int\dd
 E|E,\alpha\rangle\langle E,\alpha|\hat{J}(x)|E,\alpha^{\prime}\rangle\langle
 E,\alpha^{\prime}|.
\]
If the initial wave packet consisting only of the waves moving in the
positive direction is assumed, then one has 
\begin{eqnarray*}
\langle\hat{N}(x)\rangle & = & 2\pi\hbar\int\dd E\left\langle |E,+\rangle\langle
 E,+|\hat{J}(x)|E,+\rangle\langle E,+|\right\rangle ,\\
\langle\hat{F}(x)\hat{N}(X)\rangle & = & 4\pi^2\hbar^2\sum_{\alpha}\int\dd
 E\left\langle |E,+\rangle\langle E,+|x\rangle\langle x|E,\alpha\rangle\langle
 E,\alpha|\hat{J}(X)|E,+\rangle\langle E,+|\right\rangle .
\end{eqnarray*}
From the condition $X>L$ it follows that
\begin{equation}
\langle\hat{N}(X)\rangle=\int\dd E\left\langle |E,+\rangle|t(E)|^2\langle E,
+|\right\rangle .\label{Nave}
\end{equation}
 For $x<0$ we obtain the following expressions for the quantities
$\tau^{\mathrm{Tun}}(x,t)$ and $\tau_{\mathrm{corr}}^{\mathrm{Tun}}(x,t)$
\label{tuntimebefore}
\begin{eqnarray}
\tau^{\mathrm{Tun}}\left(x,t\right)& = &\frac{M}{\langle\hat{
N}(X)\rangle}\int\dd E\left\langle |E,+\rangle\frac{1}{2p_E}|t(E)|^2\left(2
+r(E)\exp\left(-2\frac{i}{\hbar}p_Ex\right)\right.\right.\nonumber\\
 &  & +\left.\left.r^*(E)\exp\left(2\frac{i}{\hbar}p_Ex\right)\right)\langle E,
+|\right\rangle ,\\
\tau_{\mathrm{corr}}^{\mathrm{Tun}}\left(x,t\right)& = &\frac{M}{2\langle\hat{
N}(X)\rangle}\int\dd E\left\langle |E,+\rangle\frac{1}{
ip_E}|t(E)|^2\left(r(E)\exp\left(-2\frac{i}{
\hbar}p_Ex\right)\right.\right.\nonumber\\
 &  & -\left.\left.r^*(E)\exp\left(2\frac{i}{\hbar}p_Ex\right)\right)\langle E,
+|\right\rangle .
\end{eqnarray}
 For $x>L$ these expressions take the form \label{tuntimeafter}
\begin{eqnarray}
\tau^{\mathrm{Tun}}(x,t) & = &\frac{M}{\langle\hat{N}(X)\rangle}\int\dd
 E\left\langle |E,+\rangle\frac{1}{2p_E}|t(E)|^2\left(2-\frac{t(E)}{
t^*(E)}r^*(E)\exp\left(2\frac{i}{\hbar}p_Ex\right)\right.\right.\nonumber\\
 &  & -\left.\left.\frac{t^*(E)}{t(E)}r(E)\exp\left(-2\frac{i}{
\hbar}p_Ex\right)\right)\langle E,+|\right\rangle ,\\
\tau_{\mathrm{corr}}^{\mathrm{Tun}}(x,t) & = &\frac{M}{2\langle\hat{
N}(X)\rangle}\int\dd E\left\langle |E,+\rangle\frac{i}{p_E}|t(E)|^2\left(\frac{
t(E)}{t^*(E)}r^*(E)\exp\left(2\frac{i}{
\hbar}p_Ex\right)\right.\right.\nonumber\\
 &  & -\left.\left.\frac{t^*(E)}{t(E)}r(E)\exp\left(-2\frac{i}{
\hbar}p_Ex\right)\right)\langle E,+|\right\rangle .
\end{eqnarray}

To illustrate the obtained formulae, the $\delta$-function barrier
\[
V(x)=\Omega\delta(x)
\]
and the rectangular barrier will be used. The Gaussian incident wave packet
initially is far to the left of the barrier.

\begin{figure}
\begin{center}\includegraphics[width=0.60\textwidth]{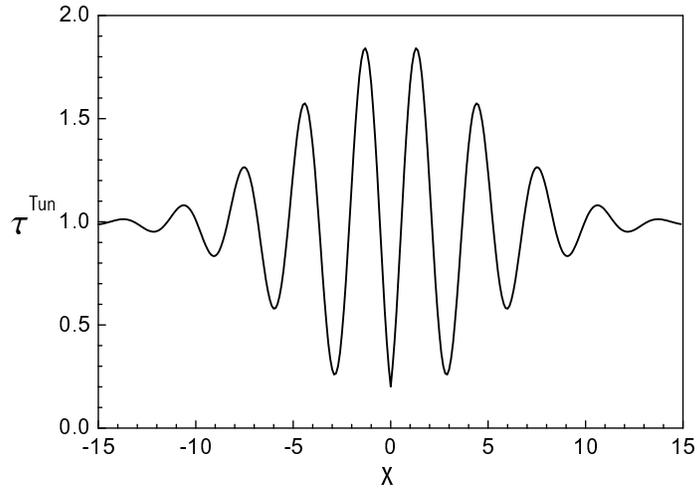}
\end{center}
\caption{The asymptotic time density for $\delta$-function barrier with the
parameter $\Omega=2$. The barrier is located at the point $x=0$. The units are
such that $\hbar=1$ and $M=1$ and the average momentum of the Gaussian wave
packet $\langle p\rangle=1$. In these units length and time are dimensionless.
The width of the wave packet in the momentum space is $\sigma=0.001$.}
\label{tundelta}
\end{figure}

\begin{figure}
\begin{center}\includegraphics[width=0.60\textwidth]{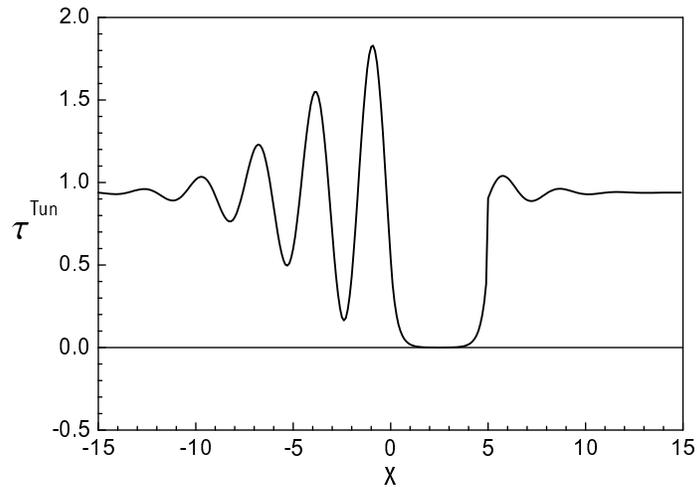}
\end{center}
\caption{The asymptotic time density for rectangular barrier. The barrier is
localized between the points $x=0$ and $x=5$ and the height of the barrier is
$V_0=2$. The used units and parameters of the initial wave packet are the same
as in Fig.~\ref{tundelta}.}
\label{tunrect}
\end{figure}

In Fig.~\ref{tundelta} and \ref{tunrect}, we see interferencelike oscillations
near the barrier. Oscillations are present not only in the front of the barrier but also
behind the barrier. When $x$ is far from the barrier the {}``time density''
tends to a value close to $1$. This is in agreement with classical mechanics
because in the chosen units the mean velocity of the particle is $1$.
Fig.~\ref{tunrect} shows additional property of {}``tunneling time density'': it
is almost zero in the barrier region. This explains the Hartmann and Fletcher
effect \cite{hartmann,fletcher}: for opaque barriers the effective tunneling
velocity is very large.

\subsection{The reflection time}

\label{secrefltime}We can easily adapt our model for the reflection too. In
doing this, one should replace the tunneling-flag operator $\hat{f}_T$ by the
reflection-flag operator 
\begin{equation}
\hat{f}_R=1-\hat{f}_T.\label{reflflag}
\end{equation}
Replacement of $\hat{f}_T$ by $\hat{f}_R$ in Eqs.~(\ref{tuntimereinf}) and
(\ref{tuntimeiminf}) gives 
\begin{equation}
\langle\tilde{f}_R(t=\infty,X)\rangle\tau^{\mathrm{Refl}}(x)=\tau^{\mathrm{
Dw}}(x)-\langle\tilde{f}_T(t=\infty,X)\rangle\tau^{\mathrm{Tun}}(x).
\label{importcond}
\end{equation}
 We see that in our model the important condition 
\begin{equation}
\tau^{\mathrm{Dw}}=T\tau^{\mathrm{Tun}}+R\tau^{\mathrm{Refl}}
\end{equation}
 where $T$ and $R$ are the transmission and reflection probabilities is satisfied
automatically.

If the wave packet consists of waves moving in the positive direction,
the density of dwell time becomes
\begin{equation}
\tau^{\mathrm{Dw}}(x,t)=2\pi\hbar\int\dd E\left\langle |E,+\rangle\langle E,
+|x\rangle\langle x|E,+\rangle\langle E,+|\right\rangle .\label{dwelltimeaprox}
\end{equation}
 For $x<0$ we have 
\begin{eqnarray}
\tau^{\mathrm{Dw}}(x,t) & = & M\int\dd E\left\langle |E,+\rangle\frac{1}{
p_E}\left(1+|r(E)|^2+r(E)\exp\left(-2\frac{i}{
\hbar}p_Ex\right)\right.\right.\nonumber\\
 &  & +\left.\left.r^*(E)\exp\left(2\frac{i}{\hbar}p_Ex\right)\right)\langle E,
+|\right\rangle
\end{eqnarray}
 and for the reflection time we obtain the {}``time density'' 
\begin{eqnarray}
\tau^{\mathrm{Refl}}(x) & = &\frac{M}{1-\langle\hat{N}(X)\rangle}\int\dd
 E\left\langle |E,+\rangle\frac{1}{p_E}\left(2|r(E)|^2\right.\right.\nonumber\\
 &  & +\frac{1}{2}\left(1+|r(E)|^2\right)r(E)\exp\left(-2\frac{i}{
\hbar}p_Ex\right)\nonumber\\
 & & +\left.\left.r^*(E)\exp\left(2\frac{i}{\hbar}p_Ex\right)\right)\langle E,
+|\right\rangle .
\end{eqnarray}
 For $x>L$ the density of the dwell time is 
\begin{equation}
\tau^{\mathrm{Dw}}(x,t)=M\int\dd E\left\langle |E,+\rangle\frac{1}{
p_E}|t(E)|^2\langle E,+|\right\rangle
\end{equation}
 and the {}``density of the reflection time'' may be expressed as
\begin{eqnarray}
\tau^{\mathrm{Refl}}(x) & = &\frac{M}{2}\int\dd E\left\langle |E,+\rangle\frac{
1}{p_E}|t(E)|^2\left(\frac{t(E)}{t^*(E)}r^*(E)\exp\left(2\frac{i}{
\hbar}p_Ex\right)\right.\right.\nonumber\\
 &  & +\left.\left.\frac{t^*(E)}{t(E)}r(E)\exp\left(-2\frac{i}{
\hbar}p_Ex\right)\right)\langle E,+|\right\rangle .
\end{eqnarray}

\begin{figure}
\begin{center}\includegraphics[width=0.60\textwidth]{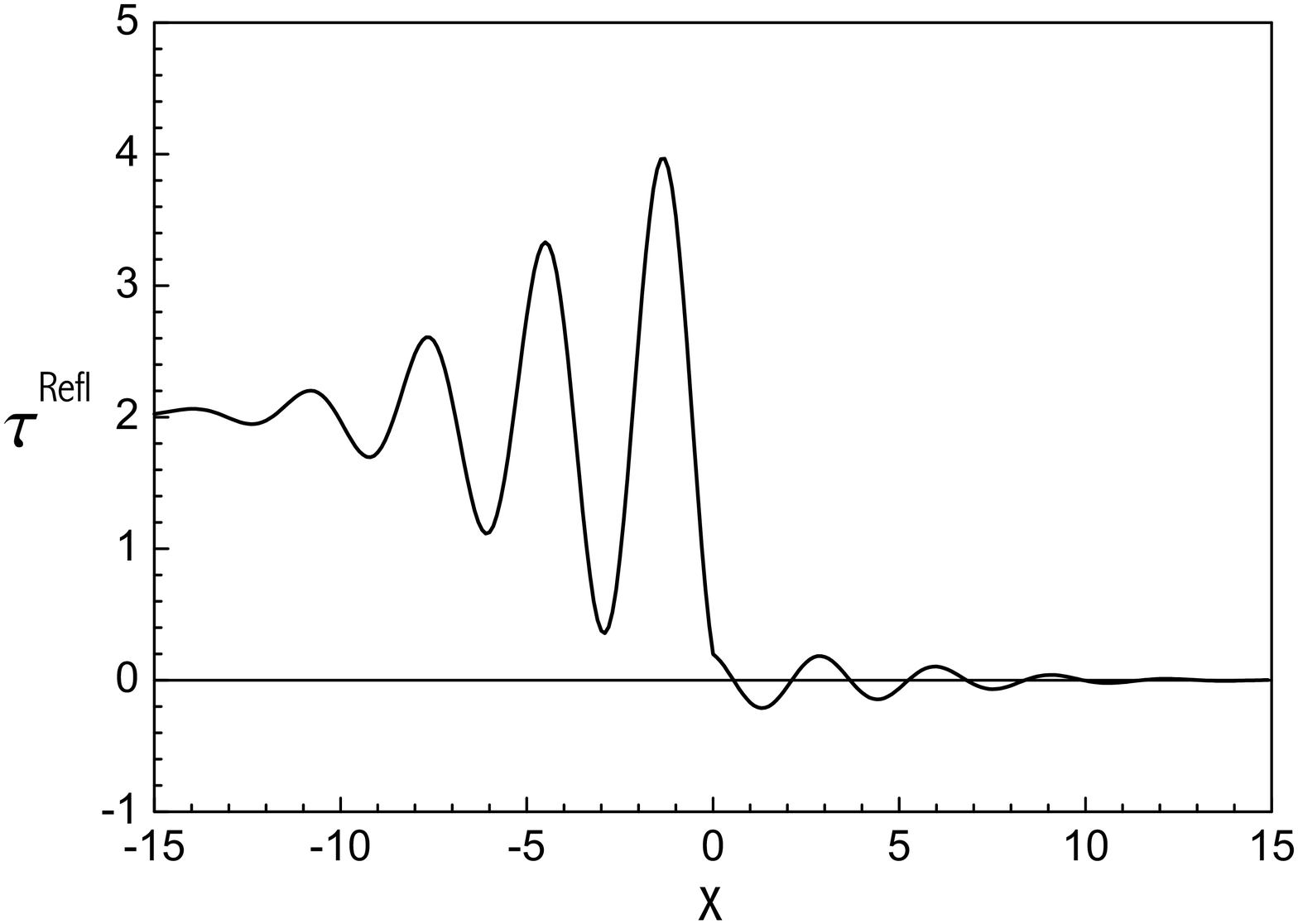}
\end{center}
\caption{Reflection time density at the same conditions as in
Fig.~\ref{tundelta}.}
\label{refldelt}
\end{figure}

\begin{figure}
\begin{center}\includegraphics[width=0.60\textwidth]{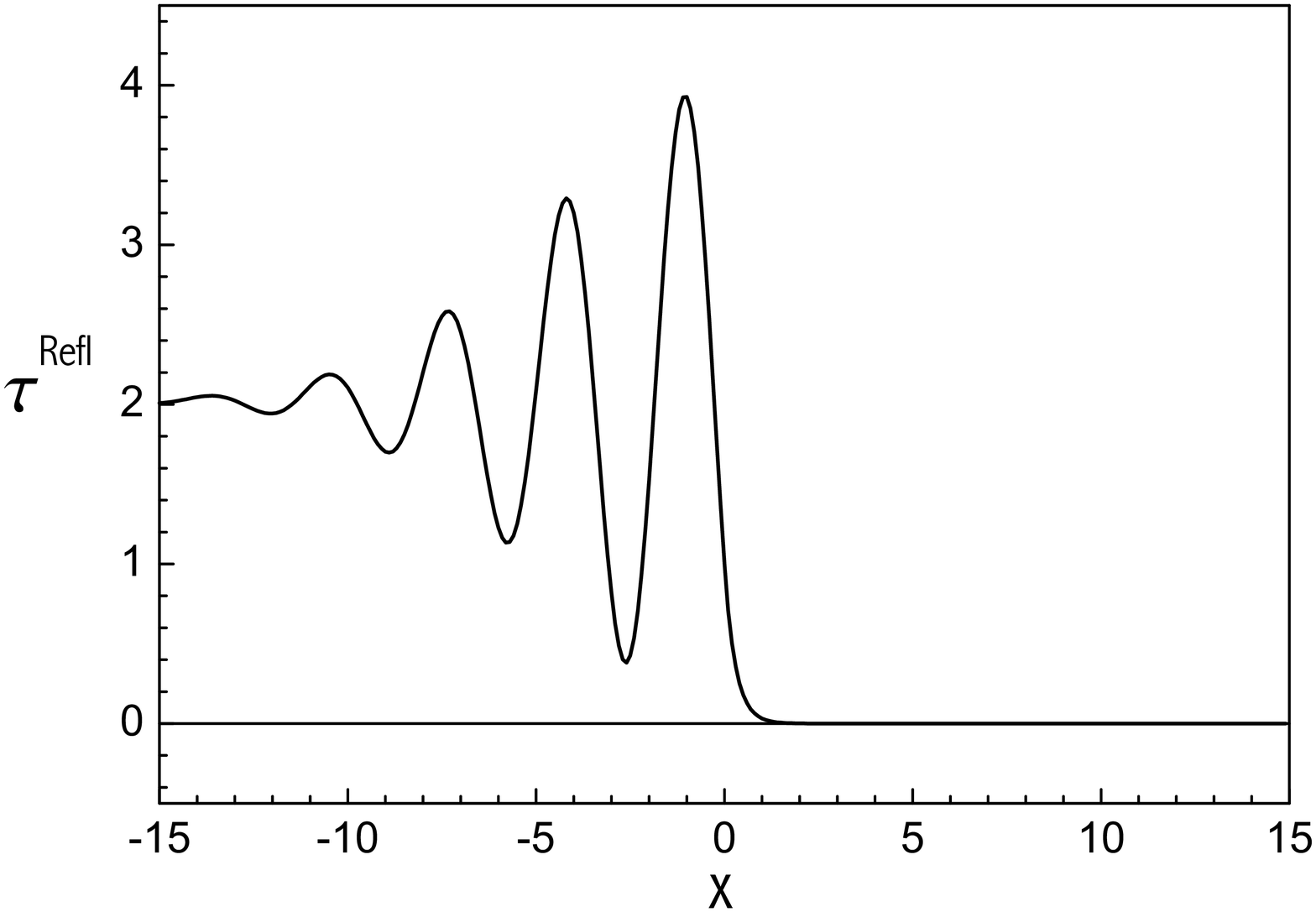}
\end{center}
\caption{Reflection time density at the same conditions as in
Fig.~\ref{tunrect}.}
\label{reflrect}
\end{figure}

\begin{figure}
\begin{center}\includegraphics[width=0.60\textwidth]{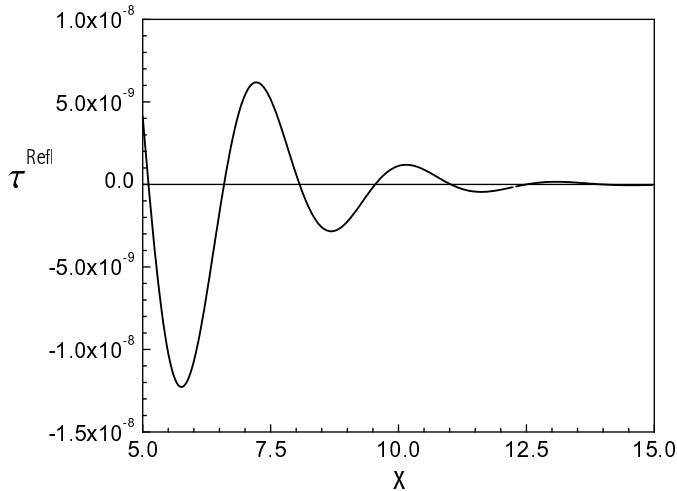}
\end{center}
\caption{Reflection time density for rectangular barrier in the region behind the
barrier. The parameters and the initial conditions are the same as in
Fig.~\ref{tunrect} }
\label{reflrect2}
\end{figure}

We will illustrate the properties of the reflection time for the same barriers
and Gaussian incident wave packet initially localized far to the left from the
barrier. In Figs.~\ref{refldelt} and \ref{reflrect}, one can see the
interference-like oscillations at both sides of the barrier. Since for the
rectangular barrier the {}``time density'' behind the
barrier is very small, this part is presented in
Fig. \ref{reflrect2}. Behind the barrier, the {}``time
density'' at certain points becomes negative. This is because the quantity
$\tau^{\mathrm{Refl}}(x)$ is not positive definite. Nonpositivity is the direct
consequence of noncommutativity of the operators in Eqs.~(\ref{tuntimereinf}) and
(\ref{tuntimeiminf}). There is nothing strange in the negativity of
$\tau^{\mathrm{Refl}}(x)$ because this quantity has no physical meaning.
Only the integral over the large region has the meaning of time. When $x$ is far
to the left from the barrier the {}``time density'' tends to a value close to
$2$ and when $x$ is far to the right from the barrier the {}``time density''
tends to $0$. This is in agreement with classical mechanics because in the
chosen units, the velocity of the particle is $1$ and the reflected particle
crosses the area before the barrier two times.

\subsection{The asymptotic time}

\label{secasympt}As mentioned above, we can determine only the time that the
tunneling particle spends in a large region containing the barrier, i.e., the
asymptotic time. In our model this time is expressed as an integral of
quantity~(\ref{tuntimereinf}) over this region. We can do the integration
explicitly.

The continuity equation yields 
\begin{equation}
\frac{\partial}{\partial t}\tilde{D}(x_D,t)+\frac{\partial}{\partial x_D}\tilde{
J}(x_D,t)=0.
\end{equation}
 The integration can be performed by parts 
\[
\int_0^t\tilde{D}(x_D,t_1)\dd t_1=t\tilde{D}(x_D,t)+\frac{\partial}{\partial
 x}\int_0^tt_1\tilde{J}(x_D,t_1)\dd t_1\,.
\]
 If the density matrix $\hat{\rho}_P(0)$ represents localized particle then
$\lim_{t\rightarrow\infty}\left(\tilde{D}(x,t)\hat{\rho}_P(0)\right)=0$.
Therefore we can write an effective equality 
\begin{equation}
\int_0^{\infty}\tilde{D}(x_D,t_1)\dd t_1=\frac{\partial}{\partial x}\int_0^{
\infty}t_1\tilde{J}(x_D,t_1)\dd t_1.
\end{equation}
 We introduce the operator 
\begin{equation}
\hat{T}(x)=\int_0^{\infty}t_1\tilde{J}(x,t_1)\dd t_1.\label{opT}
\end{equation}
 We consider the asymptotic time, i.e., the time the particle spends between
points $x_1$ and $x_2$ when $x_1\rightarrow-\infty$, $x_2\rightarrow+\infty$, 
\[
t^{\mathrm{Tun}}(x_2,x_1)=\int_{x_1}^{x_2}\tau^{\mathrm{Tun}}(x)\dd x.
\]
 After the integration we have 
\begin{equation}
t^{\mathrm{Tun}}(x_2,x_1)=t^{\mathrm{Tun}}(x_2)-t^{\mathrm{Tun}}(x_1)
\label{tuntimeasympt}
\end{equation}
 where 
\begin{equation}
t^{\mathrm{Tun}}(x)=\frac{1}{2\langle\hat{N}(x)\rangle}\left\langle\hat{
N}(x)\hat{T}(x)+\hat{T}(x)\hat{N}(x)\right\rangle .\label{asympttimeterm}
\end{equation}

If we assume that the initial wave packet is far to the left from the points
under the investigation and consists only of the waves moving in the positive
direction, then Eq.~(\ref{tuntimeasympt}) may be simplified.

In the energy representation the operator (\ref{opT}) is
\[
\hat{T}(x)=\int_{-\infty}^{\infty}t_1\dd t_1\sum_{
\alpha,\alpha'}\int\!\!\!\int\dd E\,\dd E'|E,\alpha\rangle\langle E,\alpha|\hat{
J}(x)|E',\alpha'\rangle\langle E',\alpha'|\exp\left(\frac{i}{\hbar}(E
-E')t_1\right).
\]
The integration over time yields $2i\pi\hbar^2\frac{\partial}{\partial
E^{\prime}}\delta(E-E')$ and we obtain 
\begin{eqnarray*}
\hat{T}(x) & = & -i\hbar2\pi\hbar\sum_{\alpha,\alpha'}\int\dd
 E|E,\alpha\rangle\left(\left.\frac{\partial}{\partial E'}\langle E,\alpha|\hat{
J}(x)|E',\alpha'\rangle\right|_{E'=E}\langle E,\alpha'|\right.\\
 &  & +\left.\langle E,\alpha|\hat{J}(x)|E,\alpha'\rangle\frac{\partial}{
\partial E}\langle E,\alpha'|\right),\\
\langle\hat{N}(X)\hat{T}(x)\rangle & = & -i\hbar4\pi^2\hbar^2\sum_{
\alpha}\int\dd E\langle\Psi|E,+\rangle\langle E,+|\hat{J}(X)|E,\alpha\rangle\\
 &  &\times\left(\left.\frac{\partial}{\partial E'}\langle E,\alpha|\hat{
J}(x)|E',+\rangle\right|_{E'=E}+\langle E,\alpha|\hat{J}(x)|E,+\rangle\frac{
\partial}{\partial E}\right)\langle E,+|\Psi\rangle.
\end{eqnarray*}
 Substituting expressions for the matrix elements of the probability flux
operator we obtain equation 
\begin{eqnarray*}
\langle\hat{N}(X)\hat{T}(x)\rangle & = &\int\dd E\langle\Psi|E,+\rangle
 t^*(E)\frac{\hbar}{i}\frac{\partial}{\partial E}t(E)\langle E,+|\Psi\rangle\\
 &  & +Mx\int\dd E\langle\Psi|E,+\rangle\frac{1}{p_E}|t(E)|^2\langle E,
+|\Psi\rangle\\
 &  & +i\hbar\frac{M}{2}\int\dd E\langle\Psi|E,+\rangle\frac{1}{
p_E^2}r^*(E)t^2(E)\exp\left(2\frac{i}{\hbar}p_Ex\right)\langle E,+|\Psi\rangle.
\end{eqnarray*}
When $x\rightarrow+\infty$, the last term vanishes and we have 
\begin{eqnarray}
\langle\hat{N}(X)\hat{T}(x)\rangle & = &\int\dd E\langle\Psi|E,+\rangle
 t^*(E)\frac{\hbar}{i}\frac{\partial}{\partial E}t(E)\langle E,
+|\Psi\rangle\nonumber\\
 &  & +Mx\int\dd E\langle\Psi|E,+\rangle\frac{1}{p_E}|t(E)|^2\langle E,
+|\Psi\rangle,\quad x\rightarrow+\infty.\label{positivinfnty}
\end{eqnarray}
 This expression is equal to $\langle\hat{T}(x)\rangle$,
\begin{equation}
\langle\hat{N}(X)\hat{T}(x)\rangle\rightarrow\langle\hat{T}(x)\rangle,\quad
 x\rightarrow+\infty.\label{eq:lim}
\end{equation}

When the point with coordinate $x$ is in front of the barrier, expression
(\ref{positivinfnty}) becomes
\begin{eqnarray*}
\langle\hat{N}(X)\hat{T}(x)\rangle & = & -i\hbar\int\dd E\langle\Psi|E,
+\rangle|t(E)|^2\left(\frac{i}{\hbar}\frac{M}{p_E}x\right.\\
 &  & -\left.\frac{M}{2p_E^2}r(E)\exp\left(-\frac{i}{\hbar}2p_Ex\right)+\frac{
\partial}{\partial E}\right)\langle E,+|\Psi\rangle.
\end{eqnarray*}
 When $|x|$ is large the second term vanishes and we have 
\begin{eqnarray}
\langle\hat{N}(X)\hat{T}(x)\rangle &\rightarrow & Mx\int\dd E\langle\Psi|E,
+\rangle\frac{1}{p_E}|t(E)|^2\langle E,+|\Psi\rangle\nonumber\\
 &  & +\int\dd E\langle\Psi|E,+\rangle|t(E)|^2\frac{\hbar}{i}\frac{\partial}{
\partial E}\langle E,+|\Psi\rangle.\label{negativinfnty}
\end{eqnarray}
 The imaginary part of expression~(\ref{negativinfnty}) is not zero. This means
that for determination of the asymptotic time it is insufficient to integrate
only in the region containing the barrier. For quasimonochromatic wave packets
from
Eqs.~(\ref{opT}),~(\ref{tuntimeasympt}),~(\ref{asympttimeterm}),~(\ref{positivinfnty}) and~(\ref{negativinfnty})
we obtain the limits \label{asymptmonochrom}
\begin{eqnarray}
t^{\mathrm{Tun}}(x_2,x_1) &\rightarrow & t_T^{\mathrm{Ph}}+\frac{1}{p_E}M(x_2
-x_1),\\
t_{\mathrm{corr}}^{\mathrm{Tun}}(x_2,x_1) &\rightarrow & -t_T^{\mathrm{Im}}
\end{eqnarray}
 where 
\begin{equation}
t_T^{\mathrm{Ph}}=\hbar\frac{\dd}{\dd E}(\arg t(E))\label{phasetime}
\end{equation}
 is the phase time and 
\begin{equation}
t_T^{\mathrm{Im}}=\hbar\frac{d}{
dE}\left(\ln\left|t\left(E\right)\right|\right)\label{imagtime}
\end{equation}
 is the imaginary part of the complex time.

In order to take the limit $x\rightarrow-\infty$ we have to perform more
accurate
calculations. The range of integration over time to cannot be extended
$-\infty$ because such extension corresponds to the initial wave packet being
infinitely far from the barrier. We can extend the range of the integration over
the time to $-\infty$ only in $\hat{N}(X)$. For $x<0$ we obtain
the following equation
\begin{equation}
\langle\hat{N}(X)\hat{T}(x)\rangle=\frac{1}{4\pi Mi}\int_0^{\infty}t\dd
 t\left(I_1^*(x,t)\frac{\partial}{\partial x}I_2(x,t)-I_2(x,t)\frac{\partial}{
\partial x}I_1^*(x,t)\right)
\end{equation}
 where 
\begin{eqnarray}
I_1(x,t) & = &\int\dd E\frac{1}{\sqrt{p_E}}|t(E)|^2\exp\left(\frac{i}{
\hbar}(p_Ex-Et)\right)\langle E,+|\Psi\rangle,\\
I_2(x,t) & = &\int\dd E\frac{1}{\sqrt{p_E}}\left(\exp\left(\frac{i}{
\hbar}p_Ex\right)+r(E)\exp\left(-\frac{i}{\hbar}p_Ex\right)\right)\exp\left(
-\frac{i}{\hbar}Et\right)\langle E,+|\Psi\rangle.
\end{eqnarray}
 $I_1(x,t)$ is equal to the wave function at the point $x$ and the time moment
$t$, when the propagation is in the free space and the initial wave function in
the energy representation is $|t(E)|^2\langle E,+|\Psi\rangle$. When $t\geq0$
and $x\rightarrow-\infty$, then $I_1(x,t)\rightarrow0$. That is why the initial
wave packet contains only the waves moving in the positive direction. Therefore
$\langle\hat{N}(X)\hat{T}(x)\rangle\rightarrow0$ when $x\rightarrow-\infty$.
From this analysis it follows that the region in which the asymptotic time is
well
determined has to include not only the barrier but also the initial wave packet
region.

In such a case from Eqs.~(\ref{tuntimeasympt}) and~(\ref{asympttimeterm}) we
obtain expression for the asymptotic time 
\begin{equation}
t^{\mathrm{Tun}}(x_2,x_1\rightarrow-\infty)=\frac{1}{\langle\hat{
N}(X)\rangle}\int\dd E\langle\Psi|E,+\rangle t^*(E)\left(\frac{M}{p_E}x_2
-i\hbar\frac{\partial}{\partial E}\right)t(E)\langle E,+|\Psi\rangle.
\label{mainresult}
\end{equation}
 From Eq. (\ref{eq:lim}) it follows that 
\begin{equation}
t^{\mathrm{Tun}}(x_2,x_1\rightarrow-\infty)=\frac{1}{\langle\hat{
N}(X)\rangle}\langle\hat{T}(x_2)\rangle\label{eq:fin}
\end{equation}
 where $\hat{T}(x_2)$ is defined as the probability flux integral~(\ref{opT}).
Equations (\ref{mainresult}) and~(\ref{eq:fin}) give the same value for
tunneling time as does an approach in Refs.\cite{delgadomuga,grotrovelli}

The integral of quantity $\tau_{\mathrm{corr}}^{\mathrm{Tun}}(x)$ over a large
region is zero. We have seen that it is not enough to choose the region around
the barrier---this region has to include also the initial wave packet location.
This fact will be illustrated by numerical calculations.

\begin{figure}
\begin{center}\includegraphics[width=0.60\textwidth]{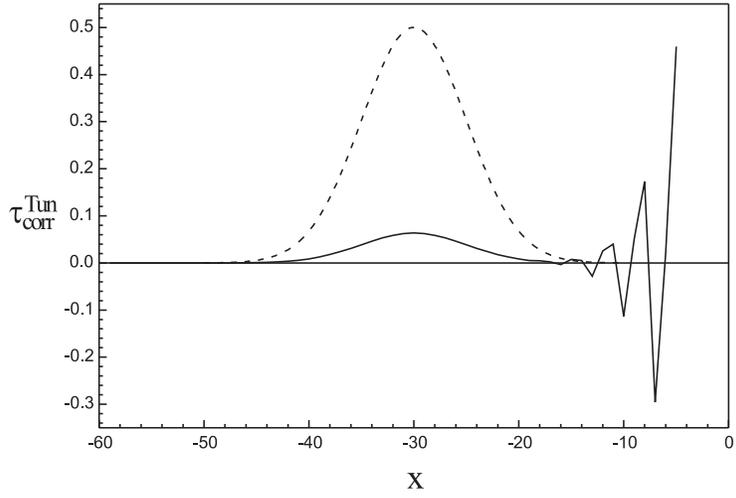}
\end{center}
\caption{The quantity $\tau_{\mathrm{corr}}^{\mathrm{Tun}}(x)$ for $\delta$
function barrier with the parameters and initial conditions as in
Fig.~\ref{tundelta}. The initial packet is shown by dashed line.}
\label{deltabarjkomutat}
\end{figure}

\begin{figure}
\begin{center}\includegraphics[width=0.60\textwidth]{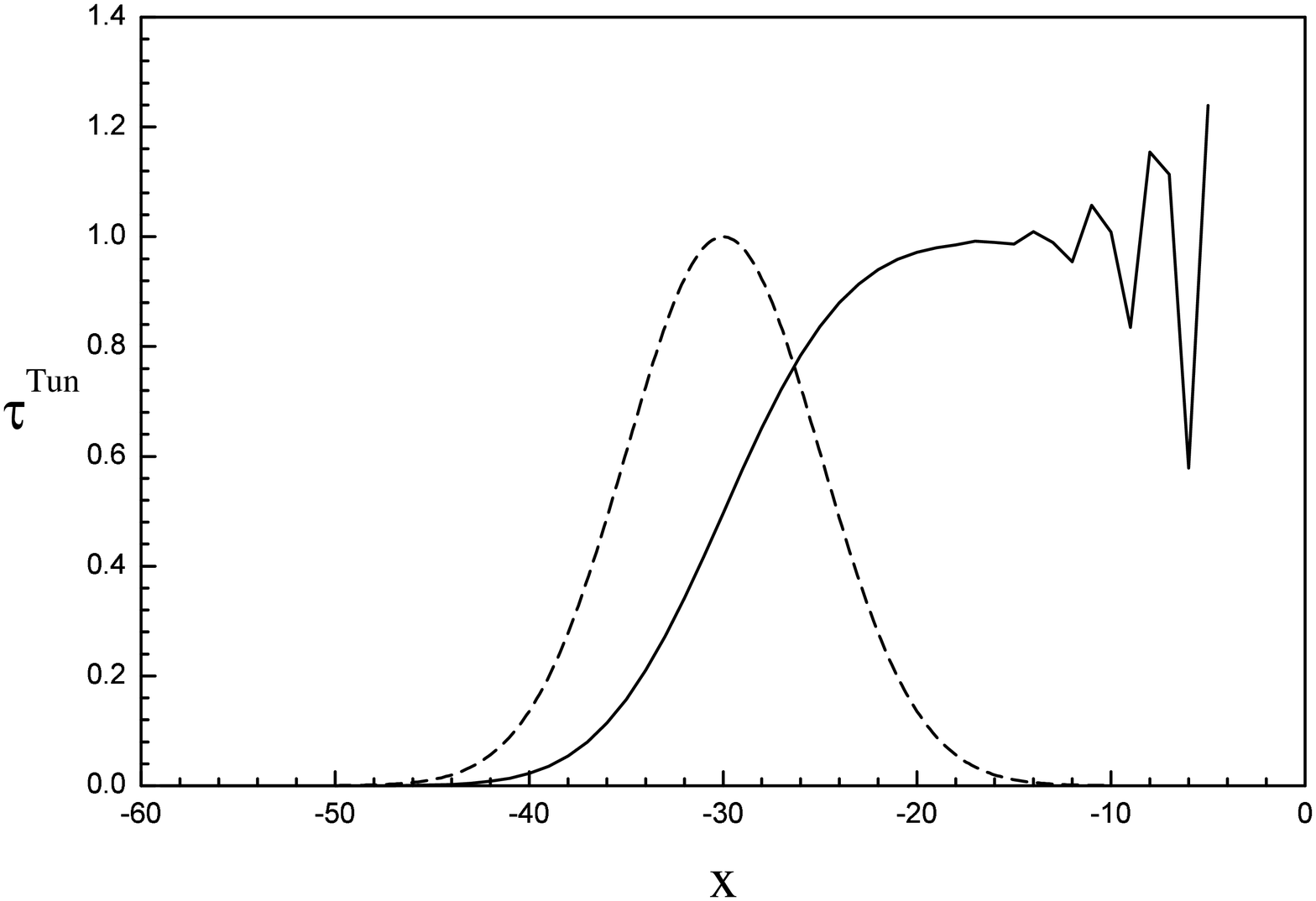}
\end{center}
\caption{Tunneling time density for the same conditions and parameters as in
Fig.~\ref{deltabarjkomutat}.}
\label{deltabarjtikslus}
\end{figure}

The quantity $\tau_{\mathrm{corr}}^{\mathrm{Tun}}(x)$ for $\delta$-function
barrier is represented in Fig. \ref{deltabarjkomutat}. We see that
$\tau_{\mathrm{corr}}^{\mathrm{Tun}}(x)$ is not equal to zero not only in the
region around the barrier but also it is not zero in the location of the initial
wave packet. For comparison, the quantity $\tau^{\mathrm{Tun}}(x)$ for the same
conditions is represented in Fig. \ref{deltabarjtikslus}.

\section{Arrival time}

\label{sec:Weak-measurement-of}The detection of the particles in time-of-flight
and coincidence experiments are common, and quantum mechanics should give a
method for the calculation of the arrival time. The arrival time distribution
may be useful in solving the tunneling time problem, as well. Therefore, the
quantum description of arrival time has attracted much attention
\cite{AharBohm,Kijowski,Muga1,Baute1,Aoki,Baute2,Baute3,Baute4,Delgado,Egusquiza,
Grot,review,Damborenea,Grubl,Kreidl,egusquizamuga}.

Aharonov and Bohm introduced the arrival time operator \cite{AharBohm}
\begin{equation}
\hat{T}_{\mathrm{AB}}=\frac{m}{2}\left((X-\hat{x})\frac{1}{\hat{p}}+\frac{1}{
\hat{p}}(X-\hat{x})\right).
\end{equation}
 By imposing several conditions (normalization, positivity, minimum variance,
and symmetry with respect to the arrival point $X$) a quantum arrival time
distribution for a free particle was obtained by Kijowski \cite{Kijowski}.
Kijowski's distribution may be associated with the positive operator valued
measure generated by the eigenstates of $\hat{T}_{\mathrm{AB}}$. However,
Kijowski's set of conditions cannot be applied in a general case
\cite{Kijowski}. Nevertheless, arrival time operators can be constructed even if
the particle is not free \cite{review,Brunetti}.

Since the mean arrival time even in classical mechanics can be infinite or the
particle may not arrive at all, it is convenient to deal not with the mean
arrival time and corresponding operator $\hat{T}$, but with the probability
distribution of the arrival times \cite{Ruseckas5}. The probability distribution
of the arrival times can be obtained from a suitable classical definition. The
noncommutativity of the operators in quantum mechanics is circumvented by using
the concept of weak measurements.

\subsection{\label{sec:classic}Arrival time in classical mechanics}

In classical mechanics the particle moves along the trajectory $H(x,p)=const$ as
$t$ increases. This allows us to work out the time of arrival at the point
$x(t)=X$, by identifying the point $(x_0,p_0)$ of the phase space where the
particle is at $t=0$, and then following the trajectory that passes by this
point, up to arrival at the point $X$. If multiple crossings are possible, one
may define a distribution of arrival times with contributions from all
crossings, when no distinction is made between first, second and $n$th arrivals.
In this article we will consider such a distribution.

We can ask whether there is a definition of the arrival time that is valid in
both classical and quantum mechanics. In our opinion, the words {}``the particle
arrives from the left at the point $X$ at the time $t$'' mean that: (i) at time
$t$ the particle was in the region $x<X$ and (ii) at time $t+\Delta t$ ($\Delta
t\rightarrow0$) the particle is found in the region $x>X$. Now we apply the
definition given by (i) and (ii) to the time of arrival in the classical case.

Since quantum mechanics deals with probabilities, it is convenient to use
probabilistic description of the classical mechanics, as well. Therefore, we
will consider an ensemble of noninteracting classical particles. The probability
density in the phase space is $\rho(x,p;t)$.

Let us denote the region $x<X$ as $\Gamma_1$ and the region $x>X$ as $\Gamma_2$.
The probability that the particle arrives from region $\Gamma_1$ to region
$\Gamma_2$ at a time between $t$ and $t+\Delta t$ is proportional to the
probability that the particle is in region $\Gamma_1$ at time $t$ and in region
$\Gamma_2$ at time $t+\Delta t$. This probability is 
\begin{equation}
\Pi_{+}(t)\Delta t=\frac{1}{N_{+}}\int_{\Omega}\dd p\,\dd x\,\rho(x,p;t),
\label{eq:p1}
\end{equation}
 where $N_{+}$ is the constant of normalization and the region of phase space
$\Omega$ has the following properties: (i) the coordinates of the points in
$\Omega$ are in the space region $\Gamma_1$ and (ii) if the phase trajectory
goes through a point of the region $\Omega$ at time $t$ then the particle at
time $t+\Delta t$ is in the space region $\Gamma_2$. Since $\Delta t$ is
infinitesimal, the change of coordinate during the time interval $\Delta t$ is
equal to $\frac{p}{m}\Delta t$. Therefore, the particle arrives from region
$\Gamma_1$ to region $\Gamma_2$ only if the momentum of the particle at the
point $X$ is positive. The phase space region $\Omega$ consists of the points
with positive momentum $p$ and with coordinates between $X-p/m\Delta t$ and $X$.
Then from Eq.~(\ref{eq:p1}) we have the probability of arrival time 
\begin{equation}
\Pi_{+}(t)\Delta t=\frac{1}{N_{+}}\int_0^{\infty}\dd p\int_{X-\frac{p}{m}\Delta
 t}^X\dd x\,\rho(x,p;t).\label{eq:p2}
\end{equation}
 Since $\Delta t$ is infinitesimal and the momentum of every particle is finite,
we can replace $x$ in Eq.~(\ref{eq:p2}) by $X$ and obtain the equality 
\begin{equation}
\Pi_{+}(t,X)=\frac{1}{N_{+}}\int_0^{\infty}\frac{p}{m}\rho(X,p;t)\dd p\,.
\label{eq:toaprob}
\end{equation}
 The obtained arrival time distribution $\Pi_{+}(t,X)$ is well known and has
appeared quite often in the literature (see, e.g., the review \cite{review} and
references therein).

The probability current in classical mechanics is 
\begin{equation}
J(x;t)=\int_{-\infty}^{+\infty}\frac{p}{m}\rho(x,p;t)\dd p.\label{eq:flux}
\end{equation}
 From Eqs.~(\ref{eq:toaprob}) and (\ref{eq:flux}) it is clear that the time of
arrival is related to the probability current. This relation, however, is not
straightforward. We can introduce the {}``positive probability current'' 
\begin{equation}
J_{+}(x;t)=\int_0^{\infty}\frac{p}{m}\rho(x,p;t)\dd p\label{eq:positflux}
\end{equation}
 and rewrite Eq.~(\ref{eq:toaprob}) as 
\begin{equation}
\Pi_{+}(t,X)=\frac{1}{N_{+}}J_{+}(X;t).\label{eq:7a}
\end{equation}
 The proposed \cite{Allcock,Bracken,Muga2} various quantum versions of $J_{+}$
even in the case of the free particle can be negative (the so-called backflow
effect). Therefore, the classical expression (\ref{eq:7a}) for the time of
arrival becomes problematic in quantum mechanics.

Similarly, for arrival from the right we obtain the probability density
\begin{equation}
\Pi_{-}(t,X)=\frac{1}{N_{-}}J_{-}(X;t),\label{eq:toaflux2}
\end{equation}
 where the negative probability current is 
\begin{equation}
J_{-}(x;t)=\int_{-\infty}^0\frac{|p|}{m}\rho(x,p;t)\dd p.
\end{equation}

We see that our definition given at the beginning of this section leads to the
proper result in classical mechanics. The conditions (i) and (ii) does not
involve the concept of the trajectories. We can try to use this definition also
in quantum mechanics.

\subsection{\label{sec:weakmeas}Weak measurement of arrival time}

The proposed definition of the arrival time probability distribution can be used
in quantum mechanics only if the determination of the region in which the
particle is does not disturb the motion of the particle. This can be achieved
using the weak measurements of Aharonov, Albert and Vaidman
\cite{Aharonov2,Aharonov1,Duck,Aharonov3,Aharonov4,Aharonov5}.

We use the weak measurement, described in Sec.~\ref{sec:concept}. The detector
interacts with the particle only in region $\Gamma_1$. As regards the operator 
$\hat{A}$ we take the projection operator $\hat{P}_1$ which projects into region
$\Gamma_1$. In analogy to Ref.~\cite{Aharonov1}, we define the {}``weak value''
of the probability of finding the particle in the region $\Gamma_1$ ,
\begin{equation}
W(1)\equiv\langle\hat{P}_1\rangle=\frac{\langle\hat{p}_q\rangle_0-\langle\hat{
p}_q\rangle}{\lambda\tau}.\label{eq:defin}
\end{equation}

In order to obtain the arrival time probability using the definition from
Sec.~\ref{sec:classic}, we measure the momenta $p_q$ of each detector after the
interaction with the particle. After time $\Delta t$ we perform the final,
postselection measurement on the particles of our ensemble and measure if the
particle is found in region $\Gamma_2$. Then we collect the outcomes $p_q$ only
for the particles found in region $\Gamma_2$.

The projection operator projecting into the region $\Gamma_2$ is $\hat{P}_2$. In
the Heisenberg representation this operator is
\begin{equation}
\tilde{P}_2(t)=\hat{U}(t)^{\dag}\hat{P}_2\hat{U}(t),
\end{equation}
where $\hat{U}$ is the evolution operator of the free particle. Taking the
operator $B$ from Sec.~\ref{sec:concept} as $\tilde{P}_2(\Delta t)$ and using
Eq.~(\ref{eq:defin}) we can introduce a weak value $W(1|2)$ of probability to
find the particle in the region $\Gamma_1$ on condition that the particle after time
$\Delta t$ is in the region $\Gamma_2$. The probability that the particle is in
region $\Gamma_1$ and after time $\Delta t$ it is in region $\Gamma_2$ then equals 
\begin{equation}
W(1,2)=W(2)W(1|2).\label{eq:18}
\end{equation}
When the measurement time $\tau$ is sufficiently small, the influence of the
Hamiltonian of the particle can be neglected. Using Eq.~(\ref{eq:sec2:x}) from
Sec.~\ref{sec:concept} we obtain
\begin{equation}
W(1,2)\approx\frac{1}{2}\langle\tilde{P}_2(\Delta t)\hat{P}_1+\hat{P}_1\tilde{
P}_2(\Delta t)\rangle+\frac{i}{\hbar}\left(\langle\hat{p}_q\rangle\langle\hat{
q}\rangle-\re\langle\hat{q}\hat{p}_q\rangle\right)\langle[\hat{P}_1,\tilde{
P}_2(\Delta t)]\rangle.\label{eq:w12}
\end{equation}

The probability $W(1,2)$ is constructed using conditions (i) and (ii) from
Sec.~\ref{sec:classic}: the weak measurement is performed to determine if the
particle is in the region $\Gamma_1$ and after time $\Delta t$ the strong
measurement determines if the particle is in the region $\Gamma_2$. Therefore,
according to Sec.~\ref{sec:classic}, the quantity $W(1,2)$ after normalization
can be considered as the weak value of the arrival time probability
distribution.

Equation (\ref{eq:w12}) consists of two terms and we accordingly can introduce two
quantities 
\begin{equation}
\Pi^{(1)}=\frac{1}{2\Delta t}\langle\hat{P}_1\tilde{P}_2(\Delta t)+\tilde{
P}_2(\Delta t)\hat{P}_1\rangle\label{eq:20}
\end{equation}
 and 
\begin{equation}
\Pi^{(2)}=\frac{1}{2i\Delta t}\langle[\hat{P}_1,\tilde{P}_2(\Delta t)]\rangle.
\end{equation}
 Then 
\begin{equation}
W(1,2)=\Pi^{(1)}\Delta t-\frac{2\Delta t}{\hbar}\left(\langle\hat{
p}_q\rangle\langle\hat{q}\rangle-\re\langle\hat{q}\hat{p}_q\rangle\right)\Pi^{
(2)}.\label{eq:prob12}
\end{equation}

If the commutator $[\hat{P}_1,\tilde{P}_2(\Delta t)]$ in
Eqs.~(\ref{eq:20})--(\ref{eq:prob12}) is not zero, then, even in the limit of
the very weak measurement, the measured value depends on the particular
detector. This fact means that in such a case we cannot obtain a definite value
for the arrival time probability. Moreover, the coefficient
$(\langle\hat{p}_q\rangle\langle\hat{q}\rangle-\re\langle\hat{q}\hat{
p}_q\rangle)$ may be zero for a specific initial state of the detector, e.g.,
for a Gaussian distribution of the coordinate $q$ and momentum $p_q$.

The quantities $W(1,2)$, $\Pi^{(1)}$ and $\Pi^{(2)}$ are real. However, it is
convenient to consider the complex quantity 
\begin{equation}
\Pi_C=\Pi^{(1)}+i\Pi^{(2)}=\frac{1}{\Delta t}\langle\hat{P}_1\tilde{P}_2(\Delta
 t)\rangle.\label{eq:complex}
\end{equation}
 We call it the {}``complex arrival probability''. We can introduce the
corresponding operator 
\begin{equation}
\hat{\Pi}_{+}=\frac{1}{\Delta t}\hat{P}_1\tilde{P}_2(\Delta t).
\end{equation}
 By analogy, the operator 
\begin{equation}
\hat{\Pi}_{-}=\frac{1}{\Delta t}\hat{P}_2\tilde{P}_1(\Delta t)
\end{equation}
 corresponds to arrival from the right.

The introduced operator $\hat{\Pi}_{+}$ has some of the properties of the classical
positive probability current. From the conditions $\hat{P}_1+\hat{P}_2=1$ and
$\tilde{P}_1(\Delta t)+\tilde{P}_2(\Delta t)=1$ we have 
\[
\hat{\Pi}_{+}-\hat{\Pi}_{-}=\frac{1}{\Delta t}(\tilde{P}_2(\Delta t)-\hat{P}_2).
\]
 In the limit $\Delta t\rightarrow0$ we obtain the probability current
$\hat{J}=\lim_{\Delta t\rightarrow0}(\hat{\Pi}_{+}-\hat{\Pi}_{-})$, as in
classical mechanics. However, the quantity $\langle\hat{\Pi}_{+}\rangle$ is
complex and the real part can be negative, in contrast to the classical quantity
$J_{+}$. The reason for this is the noncommutativity of the operators
$\hat{P}_1$ and $\tilde{P}_2(\Delta t)$. When the imaginary part is small, the
quantity $\langle\hat{\Pi}_{+}\rangle$ after normalization can be considered as
the approximate probability distribution of the arrival time.

\subsection{\label{sec:free}Arrival time probability}

The operator $\hat{\Pi}_{+}$ was obtained without specification of the
Hamiltonian of the particle and is suitable for free particles and for particles
subjected to an external potential as well. In this section we consider the
arrival time of the free particle.

The calculation of the {}``weak arrival time distribution'' $W(1,2)$ involves
the average value $\langle\hat{\Pi}_{+}\rangle$. Therefore, it is useful to have the
matrix elements of the operator $\hat{\Pi}_{+}$. It should be noted that the
matrix elements of the operator $\hat{\Pi}_{+}$, as well as the operator itself,
are only auxiliary quantities and do not have an independent meaning.

In the basis of momentum eigenstates $|p\rangle$, normalized according to the
condition $\langle p_1|p_2\rangle=2\pi\hbar\delta(p_1-p_2)$, the matrix elements
of the operator $\hat{\Pi}_{+}$ are 
\begin{eqnarray}
\langle p_1|\hat{\Pi}_{+}|p_2\rangle & = &\frac{1}{\Delta t}\langle p_1|\hat{
P}_1\hat{U}(\Delta t)^{\dag}\hat{P}_2\hat{U}(\Delta t)|p_2\rangle\nonumber\\
 & = &\frac{1}{\Delta t}\int_{-\infty}^X\dd x_1\int_X^{\infty}\dd x_2e^{-\frac{
i}{\hbar}p_1x_1}\langle x_1|\hat{U}(\Delta t)^{\dag}|x_2\rangle e^{\frac{i}{
\hbar}p_2x_2-\frac{i}{\hbar}\frac{p_2^2}{2m}\Delta t}.
\end{eqnarray}
 After performing the integration one obtains 
\begin{eqnarray}
\langle p_1|\hat{\Pi}_{+}|p_2\rangle & = &\frac{i\hbar}{2\Delta t(p_2
-p_1)}\exp\left(\frac{i}{\hbar}(p_2-p_1)X\right)\nonumber\\
 &  &\times\left(e^{\frac{i}{\hbar}\frac{\Delta t}{2m}(p_1^2-p_2^2)}\erfc\left(
-p_1\sqrt{\frac{i\Delta t}{2\hbar m}}\right)-\erfc\left(-p_2\sqrt{\frac{i\Delta
 t}{2\hbar m}}\right)\right),\label{eq:p1p2}
\end{eqnarray}
 where $\sqrt{i}=\exp(i\pi/4)$. When 
\[
\frac{1}{\hbar}\frac{\Delta t}{2m}(p_1^2-p_2^2)\ll1,\quad p_1\sqrt{\frac{\Delta
 t}{2\hbar m}}>1,\quad p_2\sqrt{\frac{\Delta t}{2\hbar m}}>1,
\]
 the matrix elements of the operator $\hat{\Pi}_{+}$ are 
\begin{equation}
\langle p_1|\hat{\Pi}_{+}|p_2\rangle\approx\frac{p_1+p_2}{2m}\exp\left(\frac{i}{
\hbar}(p_2-p_1)X\right).
\end{equation}
 This equation coincides with the expression for the matrix elements of the
probability current operator.

\begin{figure}
\begin{center}\includegraphics[width=0.60\textwidth]{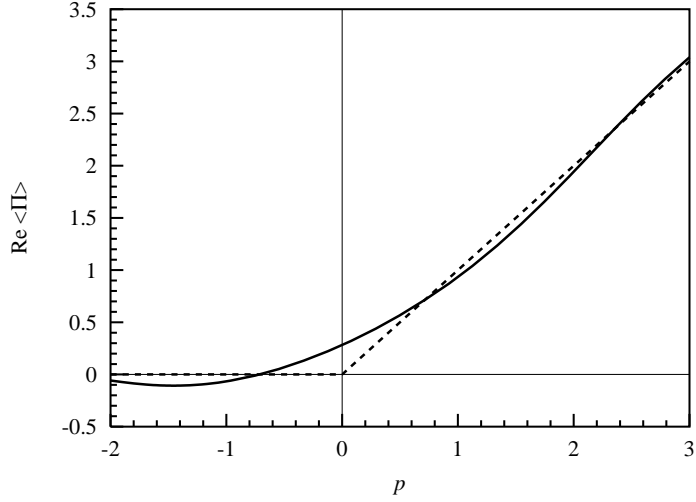}
\end{center}
\caption{The real part of the quantity $\langle p|\hat{\Pi}_{+}|p\rangle$,
according to Eq.~(\ref{eq:diagonal}). The corresponding classical positive
probability current is shown with the dashed line. The parameters used are
$\hbar=1$, $m=1$, and $\Delta t=1$. In this system of units, the momentum $p$ is
dimensionless.}
\label{fig:1}
\end{figure}

\begin{figure}
\begin{center}\includegraphics[width=0.60\textwidth]{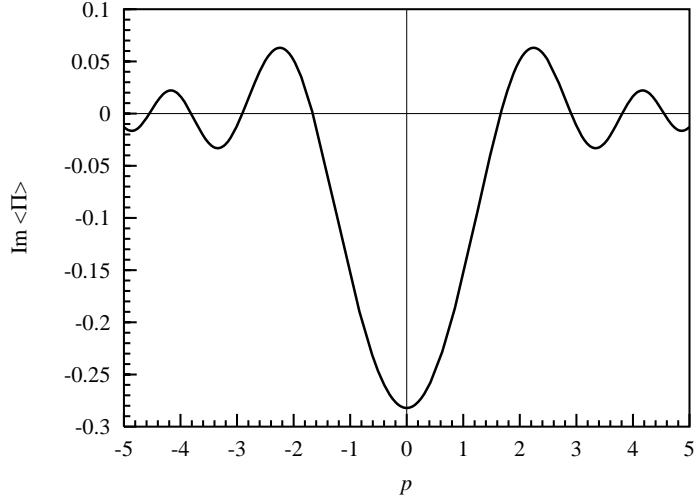}
\end{center}
\caption{The imaginary part of the quantity $\langle p|\hat{\Pi}_{+}|p\rangle$.
The parameters used are the same as in Fig. \ref{fig:1}}
\label{fig:2}
\end{figure}

>From Eq.~(\ref{eq:p1p2}) we obtain the diagonal matrix elements of the operator
$\hat{\Pi}_{+}$ ,
\begin{equation}
\langle p|\hat{\Pi}_{+}|p\rangle=\frac{p}{2m}\erfc\left(-p\sqrt{\frac{i\Delta
 t}{2\hbar m}}\right)+\frac{\hbar}{\sqrt{i2\pi\hbar m\Delta t}}e^{-\frac{i}{
\hbar}\frac{p^2}{2m}\Delta t}.\label{eq:diagonal}
\end{equation}
The real part of the quantity $\langle p|\hat{\Pi}_{+}|p\rangle$ is shown in
Fig.~\ref{fig:1} and the imaginary part in Fig.~\ref{fig:2}.

Using the asymptotic expressions for the error function $\erfc$ we obtain from
Eq.~(\ref{eq:diagonal}) that 
\[
\lim_{p\rightarrow+\infty}\langle p|\hat{\Pi}_{+}|p\rangle\rightarrow\frac{p}{m}
\]
 and $\langle p|\hat{\Pi}_{+}|p\rangle\rightarrow0$, when $p\rightarrow-\infty$,
i.e., the imaginary part tends to zero and the real part approaches the
corresponding classical value as the modulus of the momentum $|p|$ increases.
Such behaviour is evident from Figs.~\ref{fig:1} and \ref{fig:2} also.

The asymptotic expressions for function $\erfc$ are valid when the argument of
the $\erfc$ is large, i.e., $|p|\sqrt{\frac{\Delta t}{2\hbar m}}>1$ or 
\begin{equation}
\Delta t>\frac{\hbar}{E_k}.\label{eq:cond}
\end{equation}
 Here $E_k$ is the kinetic energy of the particle. 
\begin{figure}
\begin{center}\includegraphics[width=0.60\textwidth]{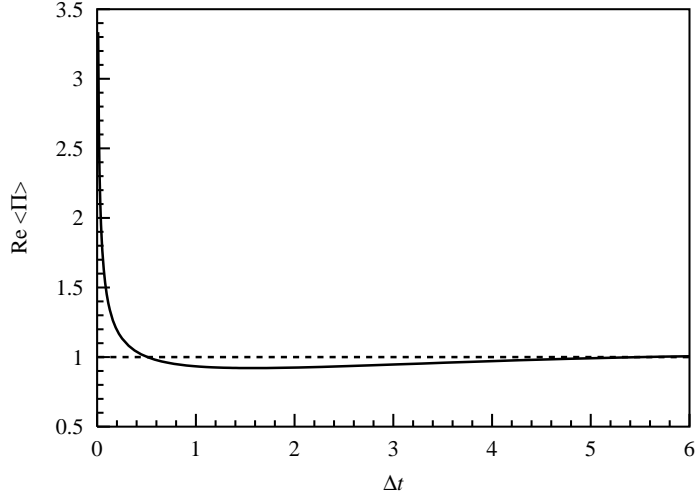}
\end{center}
\caption{The dependence of the quantity $\re\langle p|\hat{\Pi}_{+}|p\rangle$
according to Eq.~(\ref{eq:diagonal}) on the resolution time $\Delta t$. The
corresponding classical positive probability current is shown with the dashed
line. The parameters used are $\hbar=1$, $m=1$, and $p=1$. In these units, the
time $\Delta t$ is dimensionless.}
\label{fig:3}
\end{figure}
The dependence of the quantity $\re\langle
p|\hat{\Pi}_{+}|p\rangle$ on $\Delta t$ is shown in Fig.~\ref{fig:3}. For
small $\Delta t$ the quantity $\langle p|\hat{\Pi}_{+}|p\rangle$ is proportional
to $1/\sqrt{\Delta t}$. Therefore, unlike in classical mechanics, in quantum
mechanics $\Delta t$ cannot be zero. Equation (\ref{eq:cond}) imposes the lower
bound on the resolution time $\Delta t$. It follows that our model does not
permit determination of the arrival time with resolution greater than
$\hbar/E_k$. A relation similar to Eq.~(\ref{eq:cond}) based on measurement
models was obtained by Aharonov \emph{et al.\/{}} \cite{Aharonov6}. The
time-energy uncertainty relations associated with the time of arrival
distribution are also discussed in Refs.~\cite{Baute4,Brunetti2}

\section{\label{sec:concl}Summary}

The review and generalization of the theoretical analysis of the time problem
in quantum mechanics and weak measurements are presented. The tunneling time
problem is a part of this more general problem. The problem of time is solved
adapting the weak measurement theory to the measurement of time. In this model
the expression (\ref{fulltime}) for the duration, when the arbitrary observable
$\chi$ has the certain value, is obtained. This result is in agreement with the
known results for the dwell time in the tunneling time problem.

Further we consider the problem of the duration when the observable $\chi$ has
a certain value on condition that the system is in the given final state.
Our model of measurement allows us to obtain the expression (\ref{condtime}) of
this duration as well. This expression has many properties of the corresponding
classical time. However, such a duration not always has the reasonable meaning.
It is possible to obtain the duration the quantity $\chi$ has the certain value
on condition that the system is in a given final state only when the condition
(\ref{eq:poscond}) is fulfilled. In the opposite case, there is a dependence in
the outcome of the measurements on particular detector even in an ideal case
and, therefore, it is impossible to obtain the definite value of the duration.
When the condition (\ref{eq:poscond}) is not fulfilled, we introduce two
quantities (\ref{timere}) and (\ref{timeim}), characterizing the conditional
time. These quantities are useful in the case of tunneling and we suppose that
they can be useful also for other problems.

In order to investigate the tunneling time problem, we consider a procedure of
time measurement, proposed by Steinberg \cite{steinberg}. This procedure shows
clearly the consequences of noncommutativity of the operators and the
possibility of determination of the asymptotic time. Our model also reveals the
Hartmann and Fletcher effect, i.e., for opaque barriers the effective velocity
is very large because the contribution of the barrier region to the time is
almost zero. We cannot determine whether this velocity can be larger than $c$
because for this purpose one has to use a relativistic equation (e.g., the Dirac
equation).

The definition of density of one sided arrivals is proposed. This definition is
extended to quantum mechanics, using the concept of weak measurements by
Aharonov \emph{et al}
\cite{Aharonov2,Aharonov1,Duck,Aharonov3,Aharonov4,Aharonov5}\emph{.} The
proposed procedure is suitable for free particles and for the particles subjected to
an external potential, as well. It gives not only a mathematical expression for
the arrival time probability distribution but also a way of measuring the
quantity obtained. However, this procedure gives no unique expression for the
arrival time probability distribution.

In analogy with the complex tunneling time, the complex arrival time
{}``probability distribution'' is introduced (Eq.~(\ref{eq:complex})). It is
shown that the proposed approach imposes an inherent limitation,
Eq.~(\ref{eq:cond}), on the resolution time of the arrival time determination.


\end{document}